\documentclass[twocolumn,prd,nofootinbib,aps,floats,floatfix,amsmath,amssymb,secnumarabic]{revtex4} %

\usepackage{graphicx}
\usepackage[usenames,dvipsnames]{color}
\usepackage{amsmath,amssymb}
\usepackage{slashed}
\usepackage{verbatim}
\usepackage[colorlinks,citecolor=blue]{hyperref}

\def\sfrac#1#2{{\textstyle{#1\over #2}}}
\newcommand{\be}{\begin{equation}}
\newcommand{\ee}{\end{equation}}
\newcommand{\ba}{\begin{array}}
\newcommand{\ea}{\end{array}}
\newcommand{\bea}{\begin{eqnarray}}
\newcommand{\eea}{\end{eqnarray}}
\newcommand{\sss}{\scriptscriptstyle}
\newcommand{\R}{{\sss R}}

\newcommand{\B}{{\sss B}}
\newcommand{\N}{{\sss N}}
\newcommand{\nn}{\nonumber}
\renewcommand{\L}{{\sss L}}

\newcommand{\sparallel}{{\sss\parallel}}
\newcommand{\tGamma}{{\tilde \Gamma}}

\def\sfrac#1#2{{\textstyle{#1\over #2}}}

\begin{document}
\leftline{CERN-TH-2017-050}
\title{Electroweak baryogenesis from a dark sector}
\author{James M.\ Cline}
\email{jcline@physics.mcgill.ca}
\affiliation{CERN, Theoretical Physics Department, Geneva,
Switzerland}
\affiliation{Department of Physics, McGill University,
3600 Rue University, Montr\'eal, Qu\'ebec, Canada H3A 2T8}
\author{Kimmo Kainulainen}
\email{kimmo.kainulainen@jyu.fi}
\affiliation{Department of Physics, P.O.Box 35 (YFL),
FIN-40014 University of Jyv\"askyl\"a, Finland}
\affiliation{Helsinki Institute of Physics, P.O. Box 64,
             FIN-00014 University of Helsinki, Finland}
\author{David Tucker-Smith}
\email{dtuckers@williams.edu}
\affiliation{Department of Physics, Williams College, Williamstown, MA 01267}

\begin{abstract} 

Adding an extra singlet scalar $S$ to the Higgs sector can provide a 
barrier at tree level between a false vacuum with restored electroweak
symmetry and the true one.  This has been demonstrated to readily give
a strong phase transition as required for electroweak baryogenesis. We
show that with the addition of a fermionic dark matter particle $\chi$
coupling to $S$, a simple UV-complete model can realize successful 
electroweak baryogenesis.  The dark matter gets a CP asymmetry that is 
transferred to the standard model through a {\it CP portal interaction}, 
which we take to be a coupling  of $\chi$ to $\tau$ leptons and an inert 
Higgs doublet.  The CP asymmetry induced in left-handed $\tau$ leptons 
biases sphalerons to produce the baryon asymmetry. The model has promising 
discovery potential at the LHC, while robustly providing a large enough 
baryon asymmetry and correct dark matter relic density with reasonable 
values of the couplings. 

\end{abstract}
\pacs{ }
\maketitle

%%%%%%%%%%%%%%%%%%%%%%%%%%%%%%%%%%%%%%%%%%%%%%%%%%%%%%%%%%%%%%%%%%%%%%%%%%%%%%%%%%%%%%%%%%%%%%%%%%%%
%%%%%%%%%%%%%%%%%%%%%%%%%%%%%%%%%%%%%%%%%%%%%%%%%%%%%%%%%%%%%%%%%%%%%%%%%%%%%%%%%%%%%%%%%%%%%%%%%%%%
%
\section{Introduction}
%
%%%%%%%%%%%%%%%%%%%%%%%%%%%%%%%%%%%%%%%%%%%%%%%%%%%%%%%%%%%%%%%%%%%%%%%%%%%%%%%%%%%%%%%%%%%%%%%%%%%%
%%%%%%%%%%%%%%%%%%%%%%%%%%%%%%%%%%%%%%%%%%%%%%%%%%%%%%%%%%%%%%%%%%%%%%%%%%%%%%%%%%%%%%%%%%%%%%%%%%%%

Electroweak baryogensis (EWBG) is the most experimentally testable
mechanism for explaining the baryon asymmetry of the universe (BAU), and as
such it is coming under increasing pressure from LHC constraints on
new physics below the TeV scale.  The minimal supersymmetric standard
model has practically been excluded for realizing EWBG 
\cite{Liebler:2015ddv,Carena:2012np,Curtin:2012aa},  while the
remaining parameter space for two Higgs doublet models to do so is
increasingly narrow \cite{Cline:2011mm,Haarr:2016qzq,Chiang:2016vgf} 
(see however \cite{Dorsch:2016nrg}). 
The challenges are intricate since not only must
one provide new particles coupling to the Higgs field in order to make
the electroweak phase transition (EWPT) strongly first order, but also 
large CP violation to get sufficient baryon production.  Constraints
on electric dipole moments often restrict such new sources of CP
violation.  

Adding an extra scalar singlet field
to these models alleviates the tension 
\cite{Alanne:2016wtx,Huber:2006wf,Cheung:2012pg,Huang:2014ifa,Demidov:2016wcv}.
One way in which the singlet can help is by providing a tree-level
barrier that gives a robust way of making the phase
transition more strongly first order 
\cite{Choi:1993cv,Espinosa:2011ax,Espinosa:2011eu}, that has been exploited
in recent work on EWBG  
\cite{Cline:2012hg,Fairbairn:2013uta,Li:2014wia,Alanne:2014bra,
Jiang:2015cwa,Sannino:2015wka,Huang:2015bta,Xiao:2015tja,Vaskonen:2016yiu}.  Another feature of
the singlet-assisted transition is that it can lead to observable
gravitational waves 
\cite{No:2011fi,Vaskonen:2016yiu,Dorsch:2016nrg,Artymowski:2016tme,
Jaeckel:2016jlh,Katz:2016adq,
Chala:2016ykx,Huang:2016odd,Beniwal:2017eik} (see also
\cite{Dolgov:2000ht}). 

Many of the baryogenesis studies have relied upon 
dimension-5 or 6 couplings of the scalar to standard model (SM)
fermions in order to get CP-violating interactions in the bubble walls
during the phase transition.  The inverse mass scale of the
nonrenormalizable operator must be relatively small, indicating the
need for additional new physics just beyond the TeV scale.
It would be more satisfying to have an ultraviolet-complete picture
involving only renormalizable interactions.

In this paper we introduce a new realization of EWBG that meets these
criteria.  It relies upon a Majorana fermion $\chi$ that can couple to
the scalar via $S\bar\chi\gamma_5\chi$.  A $Z_2$ symmetry is imposed so that
$\chi$ can couple to standard model leptons, for example the $\tau$
lepton doublet $L_\tau$, in combination with an inert Higgs doublet
$\phi$, through the interaction $y \bar L_\tau\phi\chi$.  We dub this
a ``CP portal interaction,'' since its purpose is to transmit the CP
asymmetry between the two helicity states of $\chi$, that is generated
at the bubble wall during the first order electroweak phase
transition, to $L_\tau$ by (inverse) decays of $\phi$.  An interesting
feature of the model is that $\chi$ is a dark matter candidate if
$m_\chi < m_\phi$ (otherwise $\phi$ is the dark matter), whose relic
density is determined by the same coupling $y$ as enters into the
baryon asymmetry.  

Numerous previous works have explored possible links between
electroweak baryogenesis and dark matter~\cite{McDonald:1993ey,Profumo:2007wc,Babu:2007sm,Barger:2008jx,Ahriche:2012ei,
Gonderinger:2012rd,Fairbairn:2013xaa,Kanemura:2014cka,Jiang:2015cwa,
Lewicki:2016efe,Chala:2016ykx}.  In the case where the scalar singlet
is the dark matter candidate, it must be subdominant to the main
DM constituent if it is to provide a strong enough phase transition
for baryogenesis (while it may nevertheless be directly detectable)
\cite{Cline:2012hg}.  Unlike previous works, here the dark matter
plays an essential role in generating the baryon asymmetry rather than
strengthening the phase transition.  Moreover we show it can be
accomplished without any explicit CP violation in the dark sector,
since CP can be violated spontaneously in the DM interactions with
the scalar singlet in the bubble wall, where $S$ gets a VEV.  Since
the VEV disappears at low temperatures and CP is restored, 
the model is in this case immune to constraints from electric dipole moment searches.

We account for an issue that can be important in models with a tree
level barrier, namely the propensity for supercooling during the phase
transition, especially if it is strongly first order.  It is necessary
to construct the bubble wall solutions at temperatures below the
often considered critical temperature $T_c$, at which bubble
nucleation actually becomes faster than the Hubble rate.  We find that
moderate supercooling is required, with nucleation temperatures $T_n$
typically $20-30\%$ lower than $T_c$.  However $T_n \sim 100$ GeV, and
the new particles are not too Boltzmann suppressed to yield efficient
baryogenesis.

%%%%%%%%%%%%%%%%%%%%%%%%%%%%%%%%%%%%%%%%%%%%%%%%%%%%%%%%%%%%%%%%%%%%%%%%%%%%%%%%%%%%%%%%%%%%%%%%%%%%
%%%%%%%%%%%%%%%%%%%%%%%%%%%%%%%%%%%%%%%%%%%%%%%%%%%%%%%%%%%%%%%%%%%%%%%%%%%%%%%%%%%%%%%%%%%%%%%%%%%%
%
\section{Model and mechanism}
%
%%%%%%%%%%%%%%%%%%%%%%%%%%%%%%%%%%%%%%%%%%%%%%%%%%%%%%%%%%%%%%%%%%%%%%%%%%%%%%%%%%%%%%%%%%%%%%%%%%%%
%%%%%%%%%%%%%%%%%%%%%%%%%%%%%%%%%%%%%%%%%%%%%%%%%%%%%%%%%%%%%%%%%%%%%%%%%%%%%%%%%%%%%%%%%%%%%%%%%%%%

For simplicity, we study the phase transition in a limit of the scalar potential that has a 
$Z_2$ symmetry under $S\to -S$,
\bea
\label{pot}
	V_0 &=&  \lambda_h\left(|H|^2 -\sfrac12 v_0^2\right)^2 +
	 \sfrac14\lambda_S\left( S^2 - w_0^2\right)^2 + \sfrac12\lambda_m |H|^2
S^2\nonumber\\
\eea
Eventually one would like to consider models in which this restriction is removed, since the 
couplings of $S$ to fermions do not generically respect $S\to -S$ symmetry.  However it will 
be seen that $Z_2$ symmetry is preserved in the finite temperature scalar potential when the 
coupling of $S$ to new fermions is takes the form $i S\,\bar\chi\gamma_5\chi$, appropriate in 
the case where $S$ is interpreted as a pseudoscalar.  

In addition to the singlet scalar, we introduce a singlet Majorana fermion $\chi$ 
and an inert doublet $\phi$  
with couplings
\bea
	\sfrac12\bar \chi\left((\eta P_\R + \eta^* P_\L) S +
m_\chi\right) \chi + y \bar L_\tau \phi P_\R\chi + {\rm h.c.} \,.
\label{portal}
\eea
The hypercharge of $\phi$ is equal to that of left-handed SM leptons.
For simplicity we have coupled $\phi\chi$ only to the $\tau$ lepton doublet.
In general it could be a linear combination of lepton flavors, but with
a framework like minimal flavor violation in mind, this could be a good 
approximation to the case where the coupling to $L_\tau$ dominates.
In such a more complete setup, constraints from lepton-flavor
violating decays such as $\mu\to e\gamma$ induced at one loop by 
$\phi$ exchange will give limits on the additional couplings.

The mass $m_\chi$ is taken to be real and the coupling $\eta$
complex,  so that in general there is CP violation via their relative
phase.   The field-dependent mass has a spatially varying
phase $\theta(z)$ in the bubble walls of the phase transition, if
$S(z)$ has a nontrivial profile:
\be
\theta = {\rm arg}(\eta S + m_\chi) = 
\tan^{-1}\left(|\eta| S\over m_\chi\right)
\ee
where in the last expression we have assumed maximal phase,
$\eta = i|\eta|$, which simplifies the effective scalar potential
to be discussed below.  The coordinate $z$ denotes distance transverse to the wall, with
$z<0$ corresponding to the interior where electroweak symmetry is
broken, and $z>0$ the symmetric phase outside of the bubble.

We will show that
a baryon asymmetry can be efficiently generated by a CP asymmetry
between the two helicity components of $\chi$ that is produced at the bubble
wall and subsequently gets transmitted to $L_\tau$ leptons
by the inverse decay $\chi L_\tau\to \phi$.  
Sphalerons are sourced by the $L_\tau$ chemical potential to produce the baryon
asymmetry.  

\begin{figure}[t]
\hspace{-0.4cm}
\centerline{\includegraphics[width=0.65\hsize]{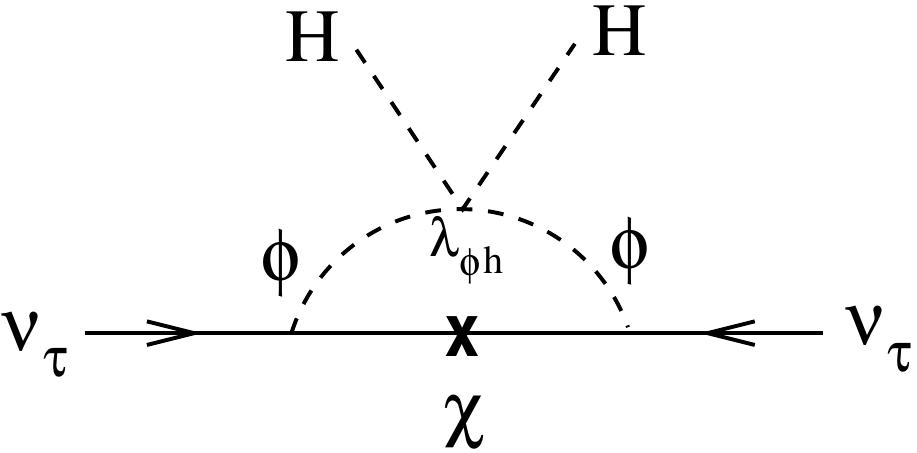}}
\caption{Loop contribution to Weinberg operator $\sim (\lambda_{\phi
h}y^2 m_\chi/16\pi^2 m_\phi^2)(L_\tau H)^2$
constrained by $\nu_\tau$ mass.}
\label{numass}
\end{figure}

For simplicity we have neglected the possible couplings of $\phi$ to
the SM Higgs doublet in the scalar potential.  
In particular, the
interaction $\lambda_{\phi h}(\phi^\dagger H)^2$ violates lepton number,  since it is
necessary to assign $\phi$ lepton number 1 in (\ref{portal}),
and consequently generates an unacceptably large Majorana mass for 
$\nu_\tau$ at one loop, as shown in fig.\ \ref{numass}.  We impose lepton number as an additional
approximate symmetry to forbid such a coupling.

%%%%%%%%%%%%%%%%%%%%%%%%%%%%%%%%%%%%%%%%%%%%%%%%%%%%%%%%%%%%%%%%%%%%%%%%%%%%%%%%%%%%%%%%%%%%%%%%%%%%
%%%%%%%%%%%%%%%%%%%%%%%%%%%%%%%%%%%%%%%%%%%%%%%%%%%%%%%%%%%%%%%%%%%%%%%%%%%%%%%%%%%%%%%%%%%%%%%%%%%%
%
\section{Effective potential and phase transition}
\label{effpot}
%
%%%%%%%%%%%%%%%%%%%%%%%%%%%%%%%%%%%%%%%%%%%%%%%%%%%%%%%%%%%%%%%%%%%%%%%%%%%%%%%%%%%%%%%%%%%%%%%%%%%%
%%%%%%%%%%%%%%%%%%%%%%%%%%%%%%%%%%%%%%%%%%%%%%%%%%%%%%%%%%%%%%%%%%%%%%%%%%%%%%%%%%%%%%%%%%%%%%%%%%%%

We follow refs.\ \cite{Espinosa:2011ax,Espinosa:2011eu}, starting from the tree-level
potential for the Higgs doublet $H$ and real singlet $S$ (\ref{pot}).
Parameters can be chosen such that the $Z_2$ symmetry
under $S\to -S$ breaks spontaneously
at high temperatures, giving $S$ a VEV (with $H=0$) in the electroweak symmetric vacuum,
while the true vacuum is along the $H$ axis at $T=0$.  The finite-temperature
effective potential for the real fields $H=h/\sqrt{2}$ and $S$ can be written in the form
\bea
	V &=& {\lambda_h\over 4}\left(h^2-v_c^2 +{v_c^2\over w_c^2}S^2\right)^2
	+ {\kappa\over 4}S^2 h^2\\
	&+& \sfrac12(T^2 - T_c^2)(c_h h^2 + c_s S^2) 
	+ \sfrac{1}{12}T^2\,{\rm Re}(\eta\, m_\chi) S\nn
\label{Veff}
\eea
where the parameter $w_0$ has been traded for its counterpart $w_c$ at the critical 
temperature of the phase transition $T_c$, $v_c$ is the corresponding critical VEV of $h$, 
and the following relations hold:
\bea
	\kappa &\equiv& \lambda_m -2\lambda_h{v_c^2\over w_c^2}\\
	T_c^2 &=& {\lambda_h\over c_h}\left(v_0^2-v_c^2\right)
\label{Tceq}
\eea
Here the coefficients $c_h$ and $c_s$ encode the $O(T^2)$ corrections to the 
masses of $h$ and $S$, and are given in terms of the gauge and other couplings by
\bea	
	c_h &=& \sfrac{1}{48}\left(9g^2+3g'^2+12y_t^2+24\lambda_h
	 + 2\lambda_m\right)\nonumber\\
	c_s &=& \sfrac{1}{12}\left(3\lambda_h{v_c^4\over w_c^4} 
	+ 2\lambda_m  + |\eta|^2\right)
\eea
where we ignored all SM Yukawa couplings apart from that of the top
quark, as well as possible couplings of the inert doublet $\phi$ to the SM Higgs.
The zero-temperature masses are given by
\bea
	m_h^2 &=& 2\lambda_h v_0^2,\\
	 m_s^2 &=& \sfrac12\kappa v_0^2 + \lambda_h(v_0^2-v_c^2)\left({v_c^2\over w_c^2}-
	{c_s\over c_h}\right)
\eea

The actual bubble nucleation temperature $T_n$ is lower than the 
critical temperature $T_c$; it is determined by
the Euclidean action $S_3$ of the bubble solution,
\be 
	S_3= 4\pi\int_0^\infty dr\, r^2\left(\sfrac12 (h'^2 + s'^2) + V(h,s)-V(0,w_T)\right)
\label{action}
\ee
($V(0,w_T)$ being the value of the potential in the false minimum
and prime denoting $d/dr$) through the relation
\be
   \exp(-S_3/T_n) = {3\over 4\pi} \left(H(T_n)\over T_n\right)^4 \left(2\pi T_n\over S_3\right)^{3/2}
\label{nuc}
\ee 
where $H$ is the Hubble rate.  

To compute the bubble action $S_3$, we discretized the 
spherically symmetric equations of
motion for $(h,s)$ following from (\ref{action}) 
and solved them using relaxation, subject to the 
boundary conditions $dh/dr = ds/dr = 0$ at $r=0$ and $(h,s)$
approaching the false minimum of the potential as $r\to\infty$.
The solutions turn out to be well described by the thin-wall
approximation~\cite{Coleman:1977py}  suitably modified to 
accommodate two fields, as
we describe in appendix~\ref{twapp}.  This approximate method
is numerically much faster than the exact solution, making it
useful for scanning over models.  As a further check, we recomputed
the nucleation temperature $T_n$ and VEV $v_n$ for models from
our random scan using the
CosmoTransitions package~\cite{Wainwright:2011kj}, verifying consistency with 
the results from our own code.\footnote{Before carrying out this more
exact analysis, we initially used the 
results of ref.~\cite{Akula:2016gpl}
which presents semianalytic formulae for the bubble wall profile 
and action.  These formula are only tractable at the lowest order
in an expansion that should converge to the accurate action and
tunneling path in field space.  We found that this lowest order
approximation typically gives $S_3/T$ much larger than the true value,
by factors of 10 or more, leading to an underestimate of $T_n$ and
an overestimate of the strength of the phase transition, $v_n/T_n$.}

%
%---------------------------------------------------------------------------------------------------
%
\begin{figure}[t]
\hspace{-0.4cm}
\centerline{\includegraphics[width=0.95\hsize]{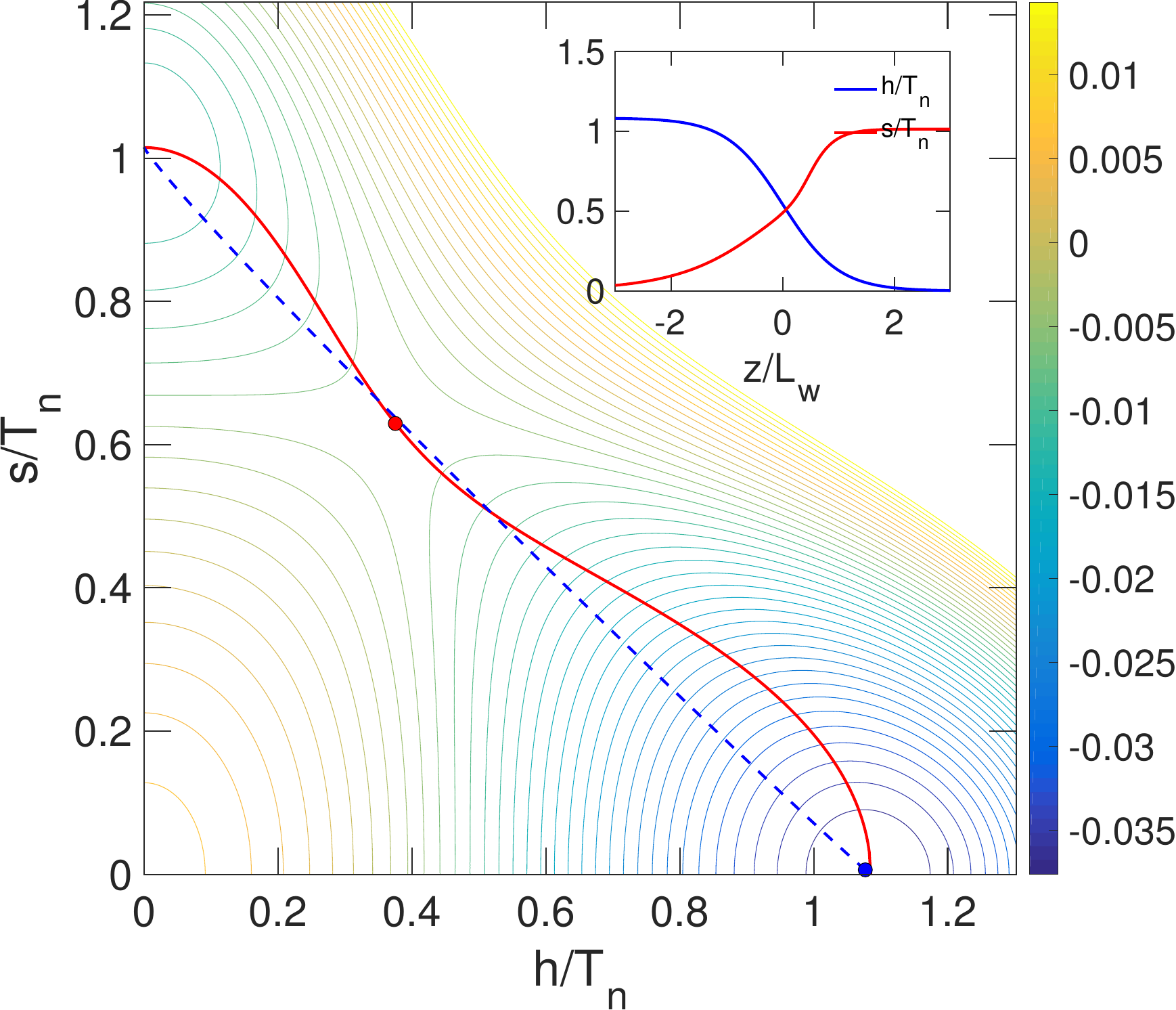}}
\caption{Example of path (for the fiducial model of table
\ref{extable}) over phase transition wall in the $(h,s)$-space (continuous red curve) with the equipotential surfaces included for $V(h,s)/T_n^4$. Red dot indicates the position of the saddle point. The amplitudes of the fields as a function of $z$ are shown in the inset. Also shown is the tunnelling path of the nucleation bubble at $T=T_n$ (blue dashed curve).}
\label{fieldpath}
\end{figure}
%
%---------------------------------------------------------------------------------------------------

Having determined the nucleation temperature and VEV, 
we demand that $v_n/T_n>1.1$ to prevent washout of the baryon asymmetry in the 
broken phase inside the bubble walls~\cite{Moore:1998swa,Fuyuto:2014yia}.  Finding the correct bubble wall 
profile is a very complicated problem that essentially depends on the friction exerted on the 
wall by the non-equilibrium plasma. Here we adopt the following simple approximation. We fix
the path in field space by minimizing the potential $V(h,s)$ along radial directions: 
$(\partial_\rho V)_\theta \equiv 0$ and setting $h(\theta)\equiv\rho(\theta)\cos\theta$ 
and $s(\theta)\equiv\rho(\theta) \sin\theta$. After this we fix the main spatial 
dependence of the profile by setting 
\bea
	h(z) &=& \sfrac12 v_n \left(1 - \tanh(z/L_w)\right) \,,
\label{wall_profile}
\eea
where the wall thickness is estimated as \cite{Espinosa:2011eu}
\vskip-0.5cm
\be
	L_w \cong \left[{2.7\over\kappa}\,{v_c^2+w_c^2\over v_c^2\,w_c^2}
	\left(1 + {\kappa\,w_c^2\over
4\lambda_h\,v_c^2}\right)\right]^{1/2} \,.
\ee
Here we use the thickness of the one-dimensional solution
corresponding to the steady-state phase of the expanding bubble wall,
rather than the thickness of the initial critical bubble determined in
\cite{Akula:2016gpl}. The former is thinner than the latter, but still
$\gtrsim 6/T$, justifying the semiclassical approach to computing the
baryon asymmetry that we employ below.  An example of the wall path in
field space, and its spatial profile, 
is shown in fig.\ \ref{fieldpath}.

%---------------------------------------------------------------------------------------------------
%
\begin{table*}[t]
\begin{tabular}{|c|c|c|c|c|c|c|c|c|c|c|c|c|c|c|c|c|}
\hline
& $m_\chi$ & $m_\phi$ & $\lambda_m$ & $y$  & $|\eta|$  & $m_s$ & $w_c$ & $w_n$ & $v_c$ & $v_n$ & $T_c$ & 
$T_n$ & $v_n/T_n$ & $L_wT_n$ & $|\eta_B|/\eta_{B,\rm obs}$ &$\Omega_{\rm dm} h^2$ \\                       
\hline
example    & 55.7 & 122.4 & 0.68 & 0.66 & 0.42 & 132.2 & 91.1 & 117.7 & 79.7 & 125.9 & 127.9 & 116.1  & 1.08 & 6.8 & 0.85 & 0.11 \\
average    & 50.5 & 117.0 & 0.68 & 0.56 & 0.40 & 129.4 & 98.8 & 117.1 & 94.2 & 133.0 & 124.7 & 113.5  & 1.18 & 5.9 & 1.32 & 0.36 \\   
std.\ dev.\ &5.3 &  10.3 & 0.12 & 0.14 & 0.17 &  12.6 & 10.2 &  12.2 & 9.4 &  10.8 &   2.3 &   4.2  & 0.15 & 0.8 & 2.13& 0.39\\   

\hline
\end{tabular}
\caption{First line: parameters for a benchmark model for succesful baryogenesis 
and observed DM abundance. Lower lines:
mean and standard deviation of parameters from 600 random models with BAU and DM relic density 
of the right order of magnitude. Units are GeV for all dimensionful parameters.  Subscript $n$ 
refers to quantities at the nucleation temperature.}
\label{extable}
\end{table*} 
%
%---------------------------------------------------------------------------------------------------

%\cite{Akula:2016gpl}
%\be
%	L_w = \ell_0\left(\left(\alpha-\frac12\right)^r + {C\over \left|
%	\alpha-\frac34\right|^s}\right)\left(v_n^2+ w_n^2\over E\right)^{1/2}
%\label{wall_width}
%\ee
%with $\ell_0 = 1.4833$, $r=18$, $C=0.4653$, $s=0.7035$.

To maintain consistency with our adoption of the $Z_2$ symmetric
tree-level potential, we can set Re($\eta\,m_\chi)=0$. This
corresponds to taking a CP-conserving coupling of $S$ to $\chi$, where
$S$ transforms as a pseudoscalar.   In a realistic treatment, there
should be some breaking of the $Z_2$ symmetry; otherwise equal and
opposite baryon asymmetries are produced from  neighboring regions of
the universe starting from false vacua with $S>0$ or $S<0$, leading to
a net vanishing asymmetry.  However a small breaking to remove the
degeneracy is sufficient to dilute away any regions in the
higher-energy vacuum by the evolution of 
domain walls, and should not change our estimate of the baryon
asymmetry in a significant way, as long as the domain walls formed when
$S$ condenses have time to move away before tunneling to the electroweak
symmetry breaking vacuum occurs.

%---------------------------------------------------------------------------------------------------
%
\begin{figure}[t]
\hspace{-0.4cm}
\centerline{\includegraphics[width=0.95\hsize]{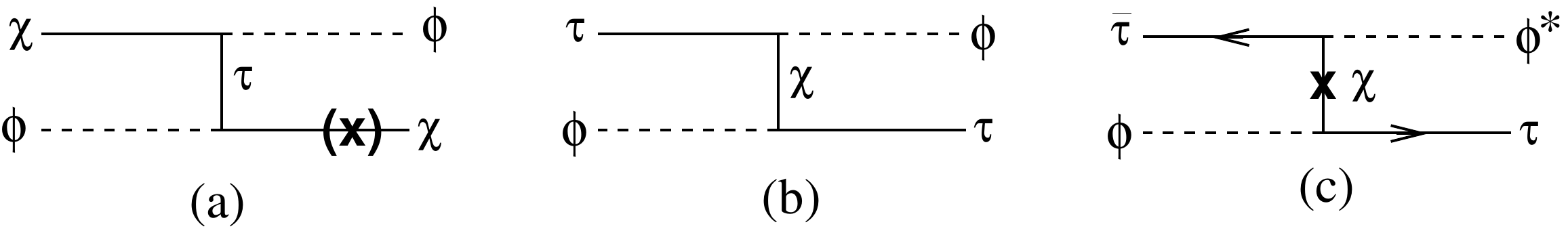}}
\caption{Dominant infrared-sensitive processes governing 
elastic and helicity flipping scattering rates.  The lines marked 
with $\mathbf{(x)}$ indicate possible helicity flips of the $\chi$ that enter into the rate 
$\Gamma_{hf}$.  These are computed in the electroweak symmetric phase in front of the wall, 
neglecting hypercharge interactions.}
\label{reactions}
\end{figure}
%
%---------------------------------------------------------------------------------------------------

%%%%%%%%%%%%%%%%%%%%%%%%%%%%%%%%%%%%%%%%%%%%%%%%%%%%%%%%%%%%%%%%%%%%%%%%%%%%%%%%%%%%%%%%%%%%%%%%%%%%
%%%%%%%%%%%%%%%%%%%%%%%%%%%%%%%%%%%%%%%%%%%%%%%%%%%%%%%%%%%%%%%%%%%%%%%%%%%%%%%%%%%%%%%%%%%%%%%%%%%%
%
\section{Baryon asymmetry}
%
%%%%%%%%%%%%%%%%%%%%%%%%%%%%%%%%%%%%%%%%%%%%%%%%%%%%%%%%%%%%%%%%%%%%%%%%%%%%%%%%%%%%%%%%%%%%%%%%%%%%
%%%%%%%%%%%%%%%%%%%%%%%%%%%%%%%%%%%%%%%%%%%%%%%%%%%%%%%%%%%%%%%%%%%%%%%%%%%%%%%%%%%%%%%%%%%%%%%%%%%%

To compute the chemical potentials that induce baryon violation, we use the
first-order diffusion equations for the chemical potential $\mu_i$ and velocity
perturbations  $u_i$ for $i = \chi,\, \phi,\, L_\tau$\footnote{$u_\L$ has dimensions of mass, and is
proportional to the actual velocity perturbation}, following the method introduced for the MSSM in 
ref.\ \cite{Cline:2000nw}, which was refined by~\cite{Fromme:2006wx}.  Here
$\mu_\chi$ is the potential for
negative minus positive helicity $\chi$'s.  In the limit $m_\chi\to 0$, the positive and negative
helicity states 
would correspond to the particle and antiparticle states of the massless fermion.  
With $m_\chi>0$, there is no distinction between particle and antiparticle, but there
is still approximate conservation of the helicities, whose damping by mass effects will
be taken into account in the Boltzmann equations.  Following the notation of ref.\ 
\cite{Fromme:2006wx}, the ensuing fluid equations can be concisely written as 
\bea
\label{diffeqs}
A_\chi \left({\mu_\chi' \atop u_\chi'}\right)
	&=& C_\chi
	-v_w{m_\chi^2}' \left({ K_{2,\chi}\mu_\chi \atop K_{6,\chi} u_\chi} \right)
	+ \left({0 \atop  S^h_\chi } \right)
\nn\\
A_i \left({\mu_i'\atop u_i'}\right)
	&=& C_i,  \quad {\rm for}\quad  i= \phi, \tau
\eea
where the coefficient matrices $A_i$ and the collision factors $C_i$ are given by
\be
A_i \equiv	\left({v_w K_{1,i}\atop -K_{4,i}}{1\atop v_w K_{5,i}}\right),
\quad
C_i \equiv \left({C^\mu_i \atop - \tGamma_{\rm el,i}\, u_i}\right),
\ee
with
\bea
\label{Cs}
C^\mu_\chi &=&  2\tGamma_{\rm hf}\, \mu_\chi  + 2\tGamma_{d}\, (\mu_\chi +c\mu_\tau-c\mu_\phi) 
\nn \\
C^\mu_\phi &=& \tGamma_{d}\, (\mu_\phi -\mu_\tau-c\mu_\chi)+ 2\tGamma_{\times,\phi}(\mu_\phi -\mu_\tau)
\nn\\
C^\mu_\tau &=& \tGamma_{d}(\mu_\tau+c\mu_\chi-\mu_\phi) + 2\tGamma_{\times,\tau}(\mu_\tau -\mu_\phi) \,.
\eea
Here primes denote $d/dz$, where $z$ is the direction transverse to the bubble
wall, $v_w$ is the wall velocity, $m_\chi = (m_\chi^2 +|\eta|^2S^2)^{1/2}$ is the magnitude of 
the field-dependent $\chi$ mass and thermal functions $K_{i,j}\equiv K_i(m_j/T)$ are defined 
in~\cite{Fromme:2006wx}. 

The reaction rates appearing in (\ref{diffeqs}) depend primarily upon
the (inverse) decay rates for $\phi\to L_\tau\chi$ and on the
scattering processes shown in fig.~\ref{reactions}.  These are the
dominant reactions because of infrared enhancement when the
intermediate particle exchanged in the $t$-channel goes on shell, as
described in appendices~\ref{elastic_rates}  and~\ref{hf_rates}.  The
tildes denote a particular normalization  of the rates (see eq.\
(\ref{eq:gtilde}) and following eq.\ (\ref{eq:Knfuns}))) that are
convenient for verifying conserved quantities; in the present case
total lepton number $n_\phi + n_{\tau}$ which is conserved by the
$\bar L_\tau\phi\chi$ interaction. $\tGamma_{\rm hf}$ is the rate of
helicity-flipping  scatterings due to the diagram of fig.\
\ref{reactions}(a) with the $\chi$ mass insertion. $\tGamma_{\times}$
is the rate of $\phi\bar{L}_\tau \to\phi^*L_\tau$  scatterings  via
mass insertion of the internal $\chi$ shown in fig.\ 
\ref{reactions}(c). $\tGamma_{{\rm el},i}$ is the
elastic scattering rate for  particle $i$ and $\tGamma_d$ is the
(inverse) decay rate $\phi\to L_\tau \chi$. Details of the 
computations of the elastic and helicity-flipping rates are given in
appendices~\ref{elastic_rates}  and~\ref{hf_rates}.  

The deviation of the coefficient $c$ from unity quantifies the probability of helicity reversal 
of $\chi$ in a decay process due to its thermal motion. It is crucial for the mechanism that $c>0$ 
since otherwise there would be no net production of $\mu_\tau$ from the (inverse) decays, as needed 
to bias the sphalerons. However as discussed in appendix~\ref{decays}, we are far from the regime 
where $c$ would be small enough to significantly suppress the baryon asymmetry. 

\begin{figure}[t]
\hspace{-0.4cm}
\centerline{\includegraphics[width=0.85\hsize]{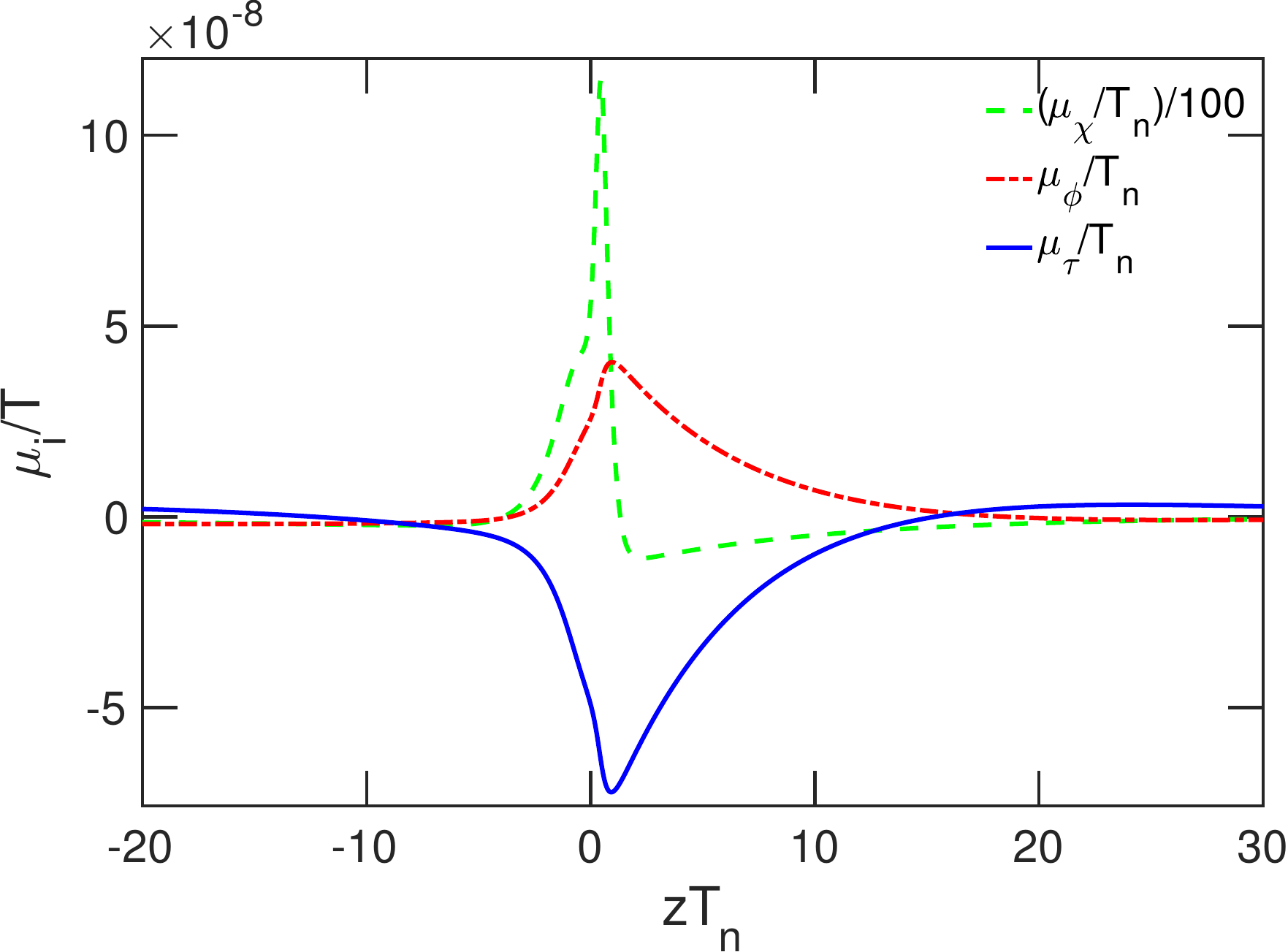}}
\caption{Solution to diffusion equations for our benchmark model
defined in Table 1.}
\label{mufig}
\end{figure}

Finally, $S^h_\chi$ is the semiclassical source, which was first derived 
in~\cite{Cline:2000nw, Kainulainen:2001cn} in the 1D-case 
and in~\cite{Kainulainen:2002th} for 3D. However, these works give the source in a particular spin-eigenstate basis, while here we
find it more convenient to work in terms of, helicity eigenstates,
defined in the wall rest frame. We derive the
correct source in the helicity basis in appendix~\ref{sec:source}. The result is,
\be
S^h_\chi  = v_w\left( -K^h_8\,(m_\chi^2\theta')' + K^h_9\,\theta' m_\chi^2 {m_\chi^2}'\right)\,,
\label{source}
\ee
where the thermal integrals $K^h_{8,9}(m_\chi/T)$ are given in appendix~\ref{sec:source}.
An example of a solution of the fluid equations (\ref{diffeqs})
for the chemical potentials is shown in fig.\ \ref{mufig}. 

%---------------------------------------------------------------------------------------------------
%
%
%---------------------------------------------------------------------------------------------------

\begin{figure}[t]
\vspace{0.35cm}
\centerline{\includegraphics[width=0.84\hsize]{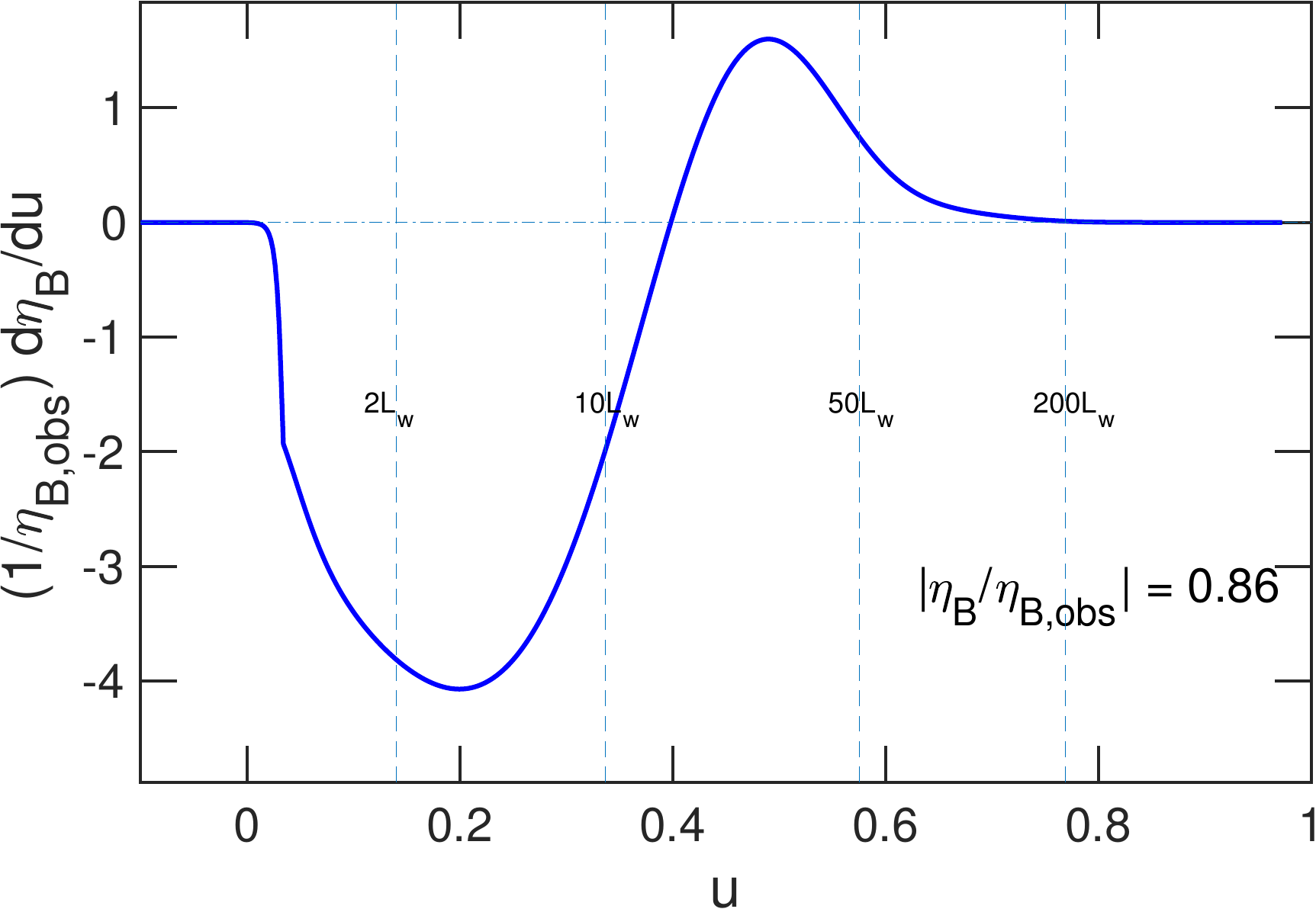}}
\caption{The differential baryon asymmetry versus distance from wall,
 ${\rm d}(\eta_B/\eta_{\rm
obs})/{\rm d}u$ for our benchmark model, where $u$ is a nonlinearly rescaled 
variable designed to optimize the grid used for 
solving the fluid equations via relaxation. The center of
the wall ($z=0$) is at $u=0$ and $z\rightarrow \pm\infty$ as
$u\rightarrow \pm 1$. Some physical distances are indicated by dashed
lines in units of the wall thickness $L_w$.}
\label{baufig}
\end{figure}

Once $\mu_\tau$ is known, the baryon-to-entropy ratio can be computed as
\be
\eta_\B = {405\,\Gamma_{\rm sph}\over
4\pi^2 \, v_w\, g_* T} \int_{-\infty}^\infty dz\, \mu_\tau\, f_{\rm sph}\, e^{-\nu z}
\label{bauint}
\ee
where
\be
f_{\rm sph}= {\rm min}\left(1,{2.4\,T\over \Gamma_{\rm sph}}e^{-40\, v(z)/T}\right)
\label{eq:fsph}
\ee
modulates the baryon violation rate to account for the local
Higgs field VEV \cite{Cline:2011mm} and $\nu = 45\,\Gamma_{\rm sph}/(4 v_w)$ accounts
for washout of the baryon asymmetry in front of the wall, if it is very slowly
moving. The sphaleron rate is $\Gamma_{\rm sph} = 1.0\times 10^{-6}\,T$.
Fig.~\ref{baufig} shows the differential asymmetry generated  around the wall
corresponding to equation (\ref{bauint}) for our benchmark model.

\begin{figure}[t]
\hspace{-0.4cm}
\centerline{\includegraphics[width=0.99\hsize]{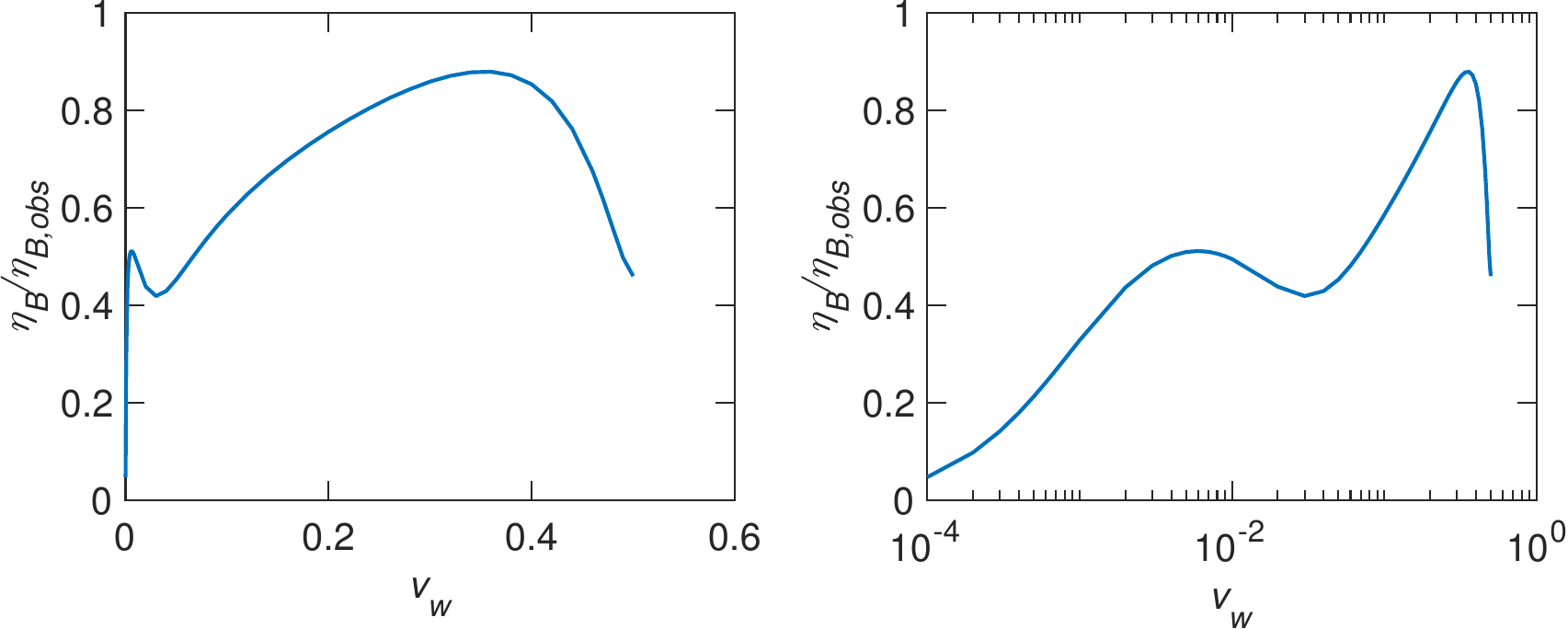}}
\caption{The baryon asymmetry as a function of the wall velocity $v_w$ for our benchmark 
model (see Table I). The right plot is the same as the left one, but using a logarithmic scale.}
\label{vwdep}
\end{figure}

The computation of the wall velocity $v_w$ is difficult 
\cite{Huber:2013kj,Konstandin:2014zta} and beyond the scope of this preliminary
study, but for typical values $v_w \sim 0.1$ our predictions are rather
insensitive to it since  the $1/v_w$ factor in (\ref{bauint}) is largely cancelled
by the $v_w$  prefactor in the source term (\ref{source}).  This is evident in
fig.\ \ref{vwdep} where we show the dependence of the baryon asymmetry on $v_w$
for the benchmark model from  Table I.   As expected, it goes to zero both for
very small  $v_w \sim 10^{-4}$ (due to the $e^{-\nu z}$ factor in eq.\ (\ref{bauint}))
and at very large $v_w$,  close to the sound velocity $1/\sqrt{3}$, where the wall
becomes a detonation, and baryogenesis is suppressed by the inability of particles
to diffuse away from the wall. Between these extremes, there is a wide plateau at
$v_w \sim 0.1-0.4$, where $\eta_B$ is only  mildly sensitive to $v_w$ and it turns
out that the situation is the same for other parameter  sets as well. This is what
motivated our assumed value for $v_w$. 

A concern for any scenario relying upon very strongly first
order transitions is that they tend to lead to faster moving walls.  Ref.\ 
\cite{Kozaczuk:2015owa}
has studied this in the context of two-field transitions such as we utilize
and shown that $v_w$ is a monotonically increasing function of $v_n/T_n$,
the strength of the phase transition at the nucleation temperature.  However
the exact value depends upon many parameters, only a few of which are covered
in ref.\ \cite{Kozaczuk:2015owa}.  Most importantly $v_w$ strongly depends upon
the friction of the wall due to its interactions with particles in the plasma.
Our model has significant new sources of such friction with sizable couplings,
through the $y\bar L_\tau\phi\chi$ and $i\eta\bar\chi \gamma_5 S\eta$
interactions.  We leave a more detailed investigation of $v_w$ in this model for
future study.

%---------------------------------------------------------------------------------------------------
%
%
%---------------------------------------------------------------------------------------------------

%---------------------------------------------------------------------------------------------------
%
\begin{figure*}[t]
\hspace{-0.4cm}
\centerline{
\includegraphics[width=0.2\hsize]{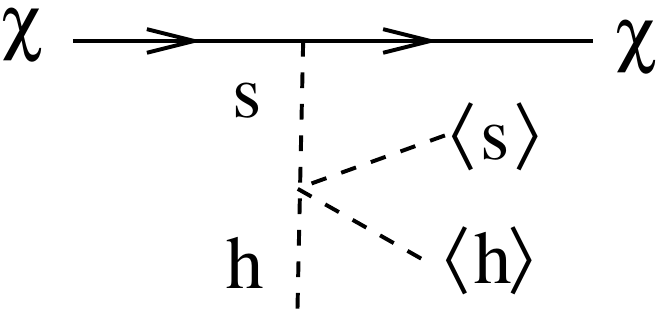}
\hfil
\includegraphics[width=0.2\hsize]{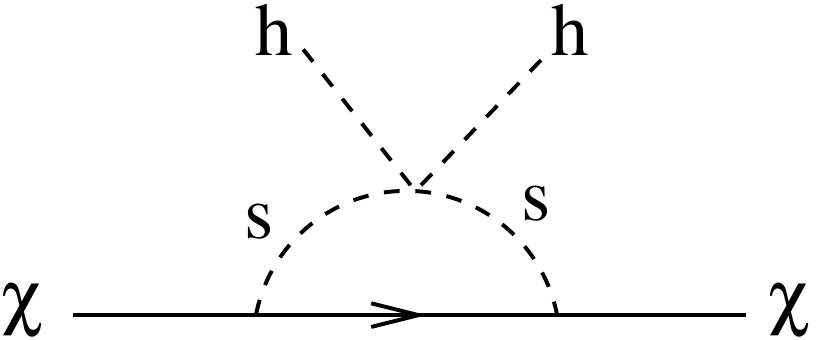}
\hfil
\includegraphics[width=0.2\hsize]{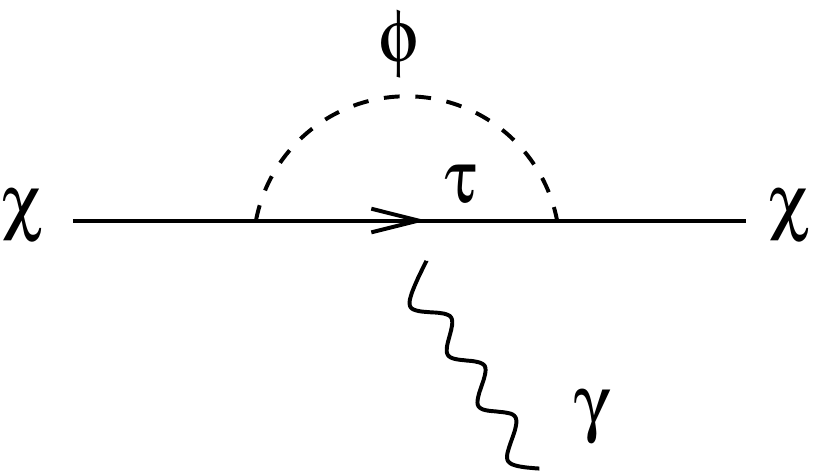}
}
\caption{Interactions contributing to dark matter-nucleon scattering.
Left: Higgs exchange in extended model where $S$ mixes with Higgs.
Middle: Higgs exchange generated by loop.
Right: Photon exchange from anapole moment.
}
\label{dd}
\end{figure*}
%

%%%%%%%%%%%%%%%%%%%%%%%%%%%%%%%%%%%%%%%%%%%%%%%%%%%%%%%%%%%%%%%%%%%%%%%%%%%%%%%%%%%%%%%%%%%%%%%%%%%%
%%%%%%%%%%%%%%%%%%%%%%%%%%%%%%%%%%%%%%%%%%%%%%%%%%%%%%%%%%%%%%%%%%%%%%%%%%%%%%%%%%%%%%%%%%%%%%%%%%%%
%
\section{Collider constraints}
%
%%%%%%%%%%%%%%%%%%%%%%%%%%%%%%%%%%%%%%%%%%%%%%%%%%%%%%%%%%%%%%%%%%%%%%%%%%%%%%%%%%%%%%%%%%%%%%%%%%%%
%%%%%%%%%%%%%%%%%%%%%%%%%%%%%%%%%%%%%%%%%%%%%%%%%%%%%%%%%%%%%%%%%%%%%%%%%%%%%%%%%%%%%%%%%%%%%%%%%%%%

The main particle physics constraint on our model is from the
Drell-Yan production of the charged Higgs bosons from the inert
doublet $\phi$, followed by their decay into $\tau$ and $\chi$, {\it
i.e.,} missing energy.  This is the same  signature as from pair
production of $\tilde\tau$ sleptons in the MSSM, so we can directly
apply such limits, since the production cross sections for
$\tilde\tau_L$ pairs is the same as for $\phi^\pm$.  ATLAS has set
limits from Run 1 which are not yet very constraining
\cite{Aad:2014yka}.  Only for $m_\chi\lesssim 20\,$GeV and $m_\phi <
130\,$GeV is the model excluded, while for $m_\chi \cong 40\,$GeV and
$m_\phi < 170\,$GeV, the allowed production cross section is less than
a factor of 2 greater than the predicted one.  CMS limits from Run 1 are comparable \cite{Khachatryan:2016trj}.
 For our benchmark
values we have chosen $m_\chi \cong 50\,$GeV, $m_\phi \cong 120\,$GeV which
should be probed during Run 2 of the LHC.
Related analyses from Run 2 \cite{ATLAS:2016uwq,ATLAS:2016ety} focus on pair production of
charginos decaying to $\tilde\tau$ rather than direct production of 
$\tilde\tau$ and so are not directly applicable
to our model. 

Stau searches at LEP have ruled out lighter values of $m_\phi \lesssim
90\,$GeV for $m_\chi\lesssim 80\,$GeV \cite{LEP}.  We avoid this
region by restricting $m_\phi>100\,$GeV in our scans.

Although we work primarily in the limit of no mixing of $S$ with the Higgs boson,
it was pointed out that this cannot be exactly true.  In the presence of small
mixing $\theta_{hs}$, even though the decay channel $h\to SS$ is kinematically
blocked, the invisible
decay $h\to \chi\bar\chi$ is possible, with rate
\be
	\Gamma_{\rm inv} = {\eta^2\theta_{hs}^2 m_h\over 16\pi}\left(1- {4 m_\chi^2\over
m_h^2}\right)^{3/2}
\ee
Demanding that the invisible branching ratio not exceed 30\% leads to the 
constraint $\theta_{hs} < 0.15$ for our benchmark parameters.

%---------------------------------------------------------------------------------------------------

%---------------------------------------------------------------------------------------------------
%
\begin{figure}[t]
\hspace{-0.4cm}
\centerline{
\includegraphics[width=0.95\hsize]{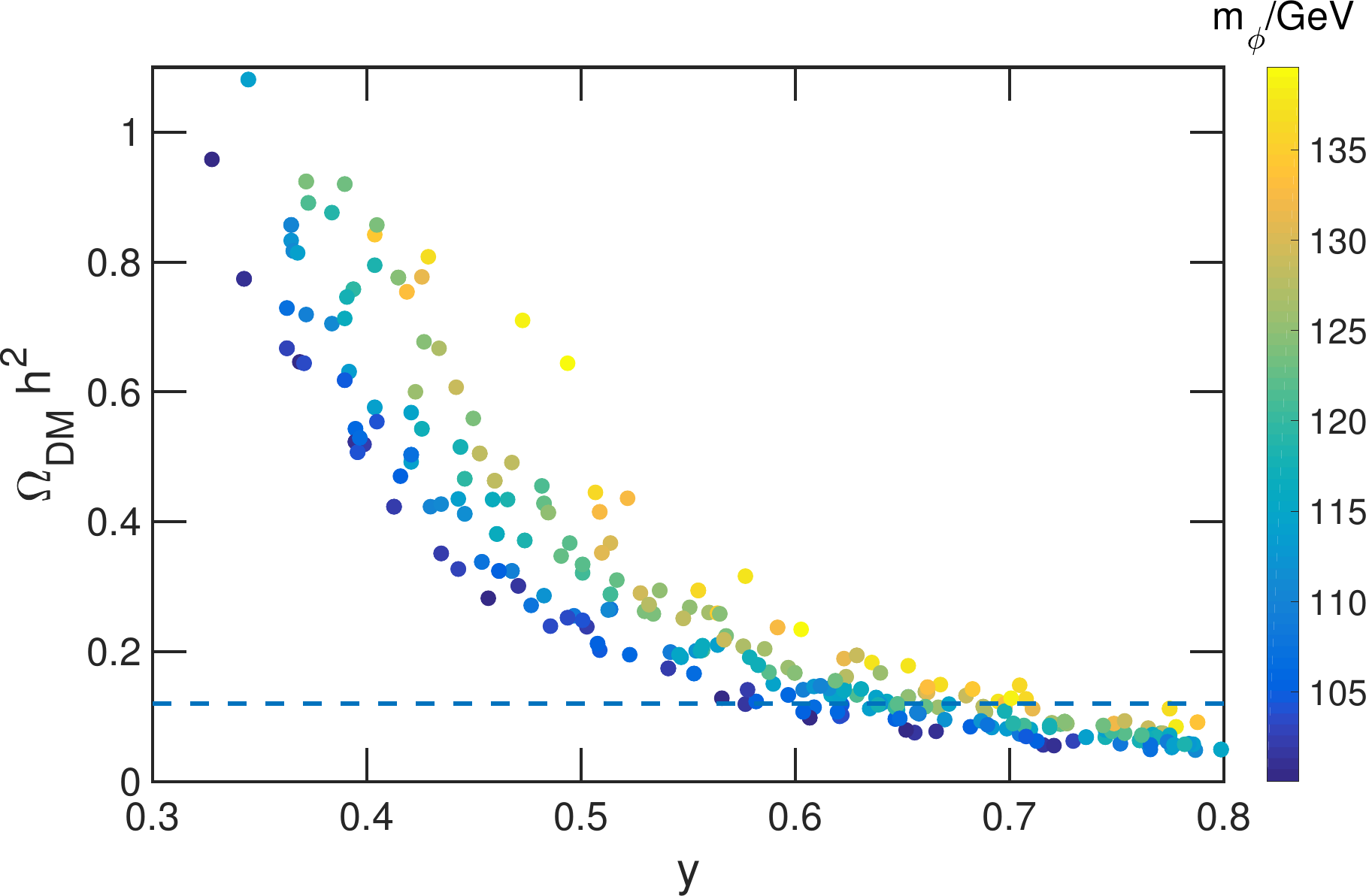}}%{Omega-y}}
\caption{Dark matter relic density versus CP-portal coupling
$y$ from scan of parameter space. Coloring of dots indicates the mass of the $\phi$ particle in GeV.}
\label{Oh2y}
\end{figure}
%
%---------------------------------------------------------------------------------------------------

%%%%%%%%%%%%%%%%%%%%%%%%%%%%%%%%%%%%%%%%%%%%%%%%%%%%%%%%%%%%%%%%%%%%%%%%%%%%%%%%%%%%%%%%%%%%%%%%%%%%
%%%%%%%%%%%%%%%%%%%%%%%%%%%%%%%%%%%%%%%%%%%%%%%%%%%%%%%%%%%%%%%%%%%%%%%%%%%%%%%%%%%%%%%%%%%%%%%%%%%%
%
\section{Dark Matter constraints}
\subsection{Relic density}
\label{relic}
%
%%%%%%%%%%%%%%%%%%%%%%%%%%%%%%%%%%%%%%%%%%%%%%%%%%%%%%%%%%%%%%%%%%%%%%%%%%%%%%%%%%%%%%%%%%%%%%%%%%%%
%%%%%%%%%%%%%%%%%%%%%%%%%%%%%%%%%%%%%%%%%%%%%%%%%%%%%%%%%%%%%%%%%%%%%%%%%%%%%%%%%%%%%%%%%%%%%%%%%%%%

Because of the $Z_2$ symmetry under which $\phi\to-\phi$ and
$\chi\to -\chi$, the lighter of the two of these particles is
a stable dark matter candidate.  Since $\phi$ would have a very
large scattering cross section on nuclei through its weak interactions,
it is preferable to assume $m_\chi < m_\phi$ so that $\chi$ is 
the dark matter.

The possible annihilation channels for $\chi$ are 
$\chi\chi\to L_\tau \bar L_\tau$ and $\chi\chi\to SS$.  Both
are $p$-wave suppressed.  We find that the respective cross sections
at lowest order in relative velocity are given by 
\bea
\label{s0tt}
	\langle\sigma v\rangle_{\tau\bar\tau}
	&=& {y^4\, m_\chi(m_\chi^4+m_\phi^4)\,T\over 
	4\pi\,(m_\chi^2 + m_\phi^2)^4}\equiv {\sigma_{0,\tau\tau}\over
	x} \\
	\langle\sigma v\rangle_{ss}
	&=& {\eta^4\, (m_\chi^2-m_s^2)^{5/2}\,T\over 
	4\pi\,(m_s^2- 2 m_\chi^2)^4} \equiv {\sigma_{0,ss}\over x} 
	\eea
The relative velocity is $v^2 = (s-4 m_\chi^2)/m_\chi^2$
and its thermal average is $\langle v^2\rangle = 6 T/m_\chi = 6/x$.
Our scans of parameter space favor $m_\chi < m_S$ so that the 
$\chi\chi\to SS$ channel is blocked, hence we focus on annihilations
to $\tau^+\tau^- + \bar\nu_\tau\nu_\tau$.

Using the analytic approximation of ref.\ \cite{Kolb:1990vq}, we find a
correlation between the relic density and $y$ shown in fig.\
\ref{Oh2y}.   Varying parameters in the ranges (\ref{ranges}),
we find that $0.6\lesssim y\lesssim 0.75$ can be compatible with
the observed value. Also, larger $m_\phi$ correlates with larger
DM-abundance, as expected from Eq.~(\ref{s0tt}).

%%%%%%%%%%%%%%%%%%%%%%%%%%%%%%%%%%%%%%%%%%%%%%%%%%%%%%%%%%%%%%%%%%%%%%%%%%%%%%%%%%%%%%%%%%%%%%%%%%%%
%%%%%%%%%%%%%%%%%%%%%%%%%%%%%%%%%%%%%%%%%%%%%%%%%%%%%%%%%%%%%%%%%%%%%%%%%%%%%%%%%%%%%%%%%%%%%%%%%%%%
%
\subsection{Direct detection}
\label{ddsec}
%
%%%%%%%%%%%%%%%%%%%%%%%%%%%%%%%%%%%%%%%%%%%%%%%%%%%%%%%%%%%%%%%%%%%%%%%%%%%%%%%%%%%%%%%%%%%%%%%%%%%%
%%%%%%%%%%%%%%%%%%%%%%%%%%%%%%%%%%%%%%%%%%%%%%%%%%%%%%%%%%%%%%%%%%%%%%%%%%%%%%%%%%%%%%%%%%%%%%%%%%%%

In the idealized limit of $S\to -S$ symmetry that we have considered,
there are no interactions of $\chi$ with nuclei at tree level.  But as
remarked previously, in a realistic model it is necessary to break
this symmetry to some extent, to prevent the universe from consisting
of cancelling domains in which the BAU has the same magnitude but
opposite signs.  The most natural way this could come about is if the 
coupling of $\chi$ to $S$ is no longer pure pseudoscalar but takes the form
\be
	S\bar\chi(\eta' + i\eta\gamma_5)\chi
\ee
since a $\chi$ loop will then induce a tadpole for $S$ proportional to $\eta'$,
leading to a VEV for $S$ at zero temperature. 
 The VEV implies mixing of
$S$ with the Higgs boson, hence interactions
with nuclei, as indicated in fig.\ \ref{dd} (left).  If the mixing angle is denoted by $\theta_{hs}\ll 1$,
the cross section on nucleons is 
\be
	\sigma_{\chi\N} =  {1\over \pi}\left(\theta_{hs}\eta' y_\N
	\mu_{\chi\N} (m_h^2-m_s^2)\over m_h^2 m_s^2\right)^2
\ee
where $\mu_{\chi\N}$ is the 
reduced mass and $y_\N \cong 0.3\, m_\N/v$ is the Higgs coupling to
nucleons with $v = 246\,$GeV \footnote{We omit the contribution proportional
to $\eta^2$ which is velocity-suppressed because of
the $\gamma_5$ in the $\bar\chi\chi S$ vertex, that consequently
gives a very small cross section $\sigma_{\chi\N}\sim 10^{-51}$cm$^2$
(taking $m_\chi\sim 50\,$GeV and $m_S\sim 100\,$GeV),
well below the current LUX bound, even for large mixing angle.}.
The recent constraint from PandaX-II \cite{Yang:2016odq} implies
\be
	\theta_{hs}\eta' < 0.04
\ee
for the benchmark model of table I (with $m_\chi = 56\,$GeV, $m_s =
109\,$GeV, suggesting
that future direct detection may be likely for reasonable values of 
the parameters in this extended version of the model.

In contrast, the loop-generated interactions shown in the middle and
right diagrams of fig.\ \ref{dd} lead to cross sections that are 
much smaller. (See ref.\ \cite{Chao:2016lqd} for analysis of a similar model.)
 The induced Higgs portal coupling has a cross 
section of order
\be
\sigma \cong {0.3^2\,\eta^4\,\lambda_m^2\, m_\chi^2\,m_N^4\over
	16^2\,\pi^5\,m_\phi^4\, m_h^4}\sim 10^{-52}{\rm\, cm^2}
\ee
In addition to the Higgs portal, $\chi$ gets an anapole moment interaction
\be
	{e\over\Lambda^2}\bar\chi\gamma_\mu\gamma_5\chi\, \partial_\nu F^{\mu\nu}
\ee
at one loop, with $\phi$ and $L_\tau$ in the loop.  This leads to a
velocity-suppressed cross section for scattering on protons \cite{Ho:2012bg},
\be
	\sigma_p = {e^4\over 2\pi\Lambda^4}\mu_{\chi p}^2 v^2
	\left(1+2{\mu_{\chi p}^2 \over m_p^2}\right)
\ee
where $\mu_{\chi p}\cong m_p$ is the reduced mass.  Estimating $\Lambda = 4\pi
m_\phi/y \sim 2\,$TeV from the loop, using our fiducial parameter values, 
we get a cross section of order $10^{-49}$cm$^2$, still far below
current sensitivities.

\subsection{Indirect detection}

Although the annihilations $\chi\bar\chi\to \tau^+\tau^-$ do not lead to
appreciable astrophysical signals, since they are $p$-wave suppressed, 
there are two associated processes that are not obviously innocuous.
The one-loop diagram with virtual $\tau$ in the loop connected to two photons 
leads to a monochromatic line from $\chi\bar\chi\to\gamma\gamma$.  Such lines
are constrained by Fermi/LAT \cite{Ackermann:2015lka}, requiring a cross section 
below $10^{-29}\,$cm$^3$/s for $m_\chi\sim 50\,$GeV, and the most cuspy assumed galactic 
dark matter density profile.  Using ref.\ \cite{Garcia-Cely:2016hsk}, 
we find that the predicted cross section is 
\be
	\langle\sigma v\rangle_{\gamma\gamma} \cong 0.2\,{\alpha^2 y^4\over 256\pi^5
m_\chi^2} \cong 4\times 10^{-30}{\rm cm^3/s}
\ee
taking values from the benchmark model, which is still below the least
conservative of the Fermi constraints.

The full cross section for $\chi\bar\chi\to \tau^+\tau^-$ includes an $s$-wave
contribution that is however helicity suppressed by $(m_\tau/m_\chi)^2$, which we
have neglected.  It was pointed out in ref.\ \cite{Bringmann:2007nk} that this suppression can be
overcome by internal brehmsstrahlung, where a photon is emitted by the charged
particle exchanged in the $t$ channel ($\phi^+$ in our case).  Using their
results, we find a cross section for emission of $\tau^+\tau_-\gamma$ 
\be
		\langle\sigma v\rangle_{\tau^+\tau_-\gamma}\cong 
	3\times 10^{-30}{\rm cm^3/s}
\ee
Since the photon is not monochromatic, we compare the prediction to constraints
on annihilation to general final states from Fermi/LAT observations of dwarf
galaxies \cite{Ackermann:2015zua}.  At $m_\chi\sim 50\,$GeV, the limits are of
order $10^{-26}{\rm cm^3/s}$ rendering this channel harmless.  The line searches
thus remain the most promising avenue for indirect discovery.

%%%%%%%%%%%%%%%%%%%%%%%%%%%%%%%%%%%%%%%%%%%%%%%%%%%%%%%%%%%%%%%%%%%%%%%%%%%%%%%%%%%%%%%%%%%%%%%%%%%%
%%%%%%%%%%%%%%%%%%%%%%%%%%%%%%%%%%%%%%%%%%%%%%%%%%%%%%%%%%%%%%%%%%%%%%%%%%%%%%%%%%%%%%%%%%%%%%%%%%%%
%
\section{Results}
%
%%%%%%%%%%%%%%%%%%%%%%%%%%%%%%%%%%%%%%%%%%%%%%%%%%%%%%%%%%%%%%%%%%%%%%%%%%%%%%%%%%%%%%%%%%%%%%%%%%%%
%%%%%%%%%%%%%%%%%%%%%%%%%%%%%%%%%%%%%%%%%%%%%%%%%%%%%%%%%%%%%%%%%%%%%%%%%%%%%%%%%%%%%%%%%%%%%%%%%%%%

We performed random scans over the model parameters to identify regions that give approximately 
the correct baryon asymmetry as well as the dark matter abundance.  
A suitable region of parameter space is given by 
\bea y
&\in&[0.3,0.8],\,\eta\in[0.1,0.9],\,\lambda_m\in [0.3,0.6], 
\nn\\
\,m_\chi&\in& [40,60],\ \ m_\phi\in[100,140],\nn\\
\log_{10}(v_c/w_c) &\in&
[-2,\,1.5],\,v_0/v_c \in[1.1,10]
\label{ranges}
\eea
In a random scan over 670,000 such models, with a flat prior on the intervals
(\ref{ranges}) we find 600 examples with a strong enough phase 
transition, and a BAU roughly within an order of magnigtude of the required value. A sample model, 
average values, and standard deviations\footnote{The resulting distributions are
not necessarily Gaussian or symmetric abound the mean values}
 of parameters are given in table \ref{extable}. We do not 
claim that this is the only region of parameter space that is viable, nor do we
attach any rigorous meaning to the statistics; rather our aim in this study 
is to demonstrate the existence of one such region, and to establish that it 
is not the result of any special fine tuning of parameters.  

%---------------------------------------------------------------------------------------------------
%
\begin{figure}[t]
\hspace{-0.4cm}
\centerline{
\includegraphics[width=0.95\hsize]{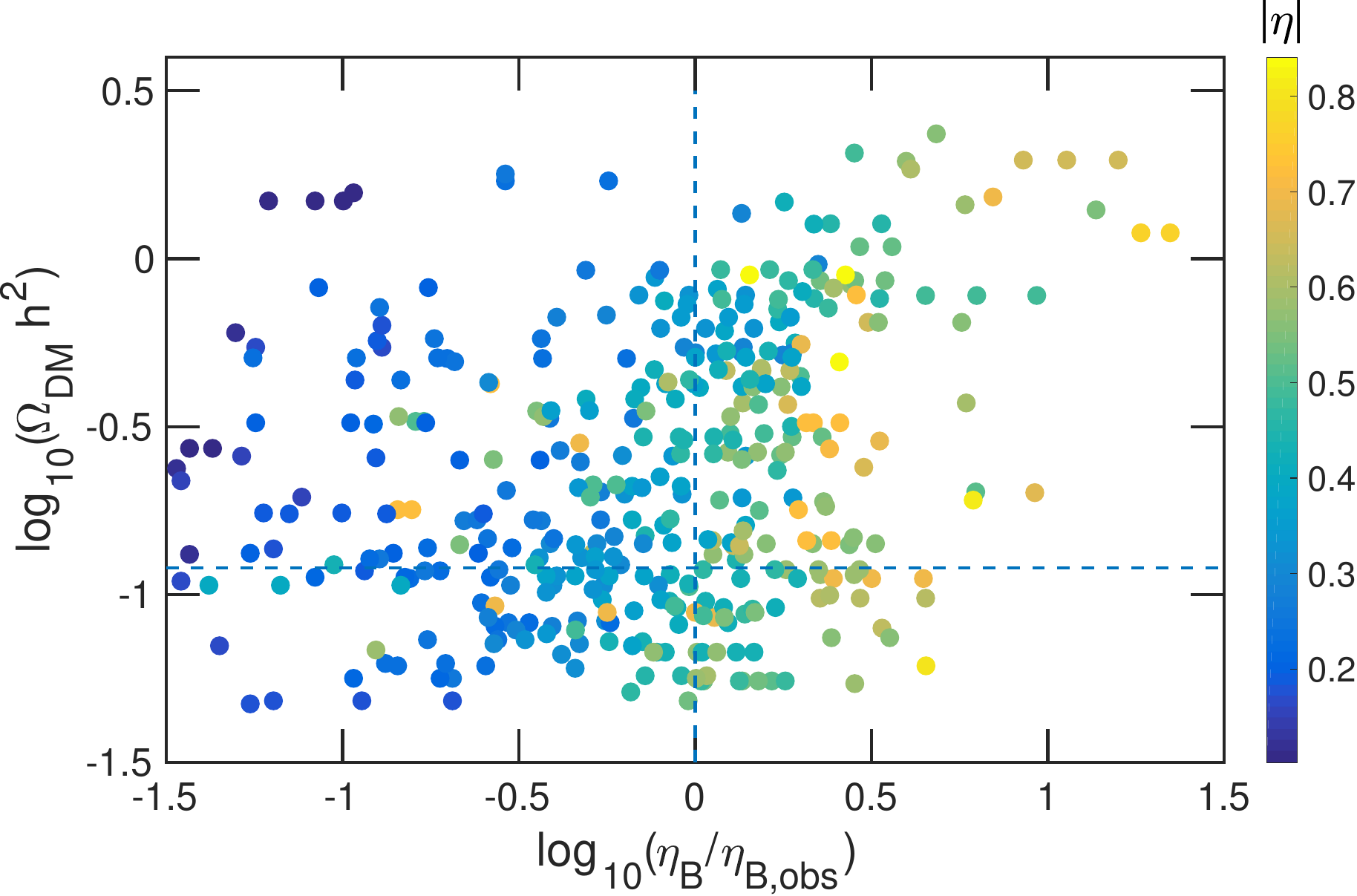}}%{Oh2-eta2}}
\caption{Relic density versus baryon asymmetry from scan of parameter space. 
Coloring of dots shows the value of the coupling $|\eta|$.}
\label{Oh2-bau}
\end{figure}
%
%---------------------------------------------------------------------------------------------------

Fig.~\ref{Oh2-bau} plots the scatter in the  plane of relic density versus baryon asymmetry, 
showing that it is easy to cover the target region of observed values. Moreover, out of the 600 
models displayed, 21 (93) have both BAU and DM within 20 (50) per cent of the observed value. The 
coloring of dots shows the absolute value of the $S\bar\chi\chi$-coupling, which controls the size
of the CP-violation in the model. As expected, the large asymmetry is strongly correlated with a 
large $|\eta|$. 

\begin{figure}[t]
\hspace{-0.4cm}
\centerline{
\includegraphics[width=\hsize]{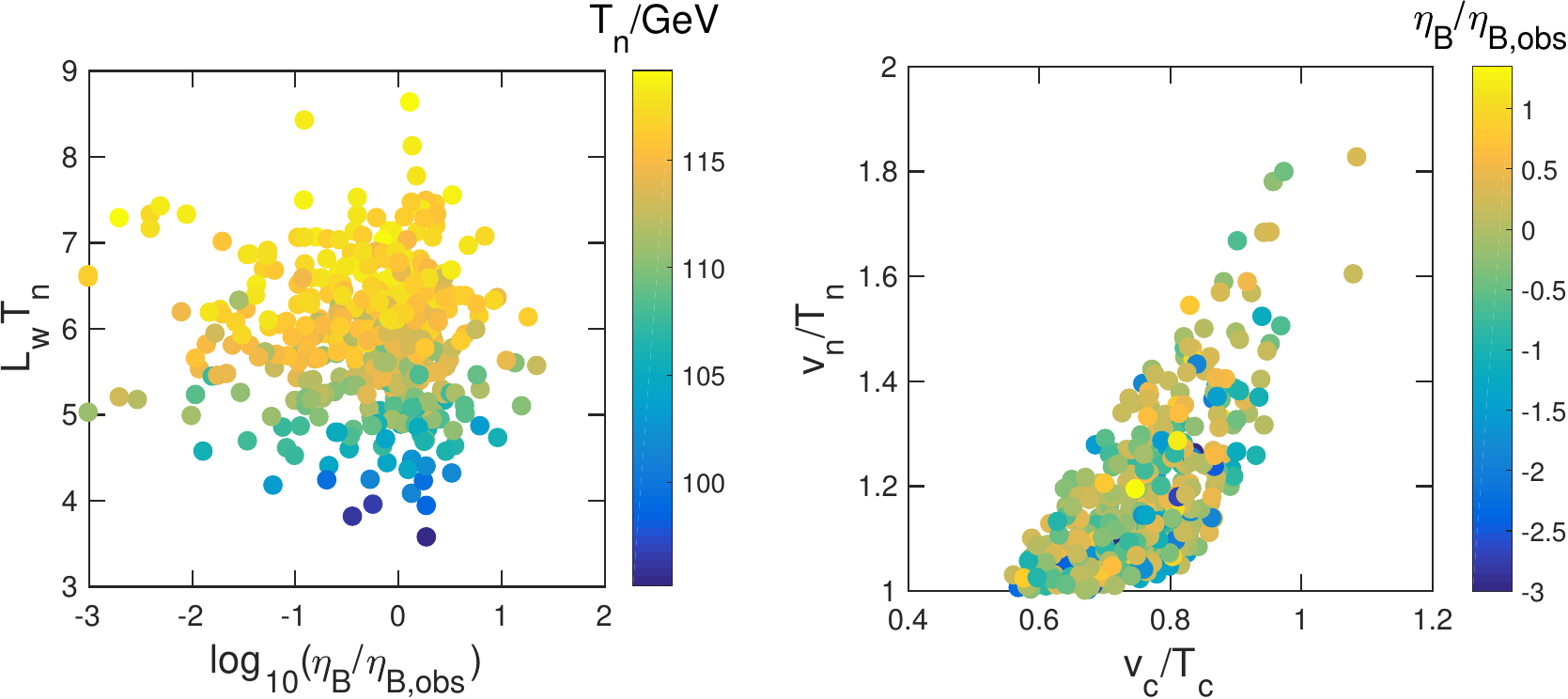}}%{Oh2-eta2}}
\caption{Left: wall width versus the baryon asymmetry. Coloring shows the nucleation temperature 
in GeV.  Right: $v_n/T_n$ as a function of $v_c/T_c$. Coloring shows the ratio of predicted and
observed baryon asymmetry.}
\label{fig:lastscat}
\end{figure}

In the left panel of Fig.~\ref{fig:lastscat} we show the correlation
between baryon  asymmetry and the wall width. There is only a slight
positive correlation between large BAU and small wall width. At any
rate all wall widths are large enough for the semiclassical method
used to  solve the fluid equations to be valid. The right panel of
Fig.~\ref{fig:lastscat} shows the correlation between  $v/T$ evaluated
at the critical and nucleation temperatures. The supercooling is never
excessive; most of the models have $v_n/T_n$ between one and two.
However some amount the supercooling is essential: {\em none} of the
models would have survived the naive sphaleron  bound $v_c/T_c > 1$.

%---------------------------------------------------------------------------------------------------
%
%
%---------------------------------------------------------------------------------------------------

%%%%%%%%%%%%%%%%%%%%%%%%%%%%%%%%%%%%%%%%%%%%%%%%%%%%%%%%%%%%%%%%%%%%%%%%%%%%%%%%%%%%%%%%%%%%%%%%%%%%
%%%%%%%%%%%%%%%%%%%%%%%%%%%%%%%%%%%%%%%%%%%%%%%%%%%%%%%%%%%%%%%%%%%%%%%%%%%%%%%%%%%%%%%%%%%%%%%%%%%%
%
\section{Conclusions}
%
%%%%%%%%%%%%%%%%%%%%%%%%%%%%%%%%%%%%%%%%%%%%%%%%%%%%%%%%%%%%%%%%%%%%%%%%%%%%%%%%%%%%%%%%%%%%%%%%%%%%
%%%%%%%%%%%%%%%%%%%%%%%%%%%%%%%%%%%%%%%%%%%%%%%%%%%%%%%%%%%%%%%%%%%%%%%%%%%%%%%%%%%%%%%%%%%%%%%%%%%%

We have demonstrated a new class of models for electroweak baryogenesis
that take advantage of a tree-level barrier facilitated by a singlet
scalar field to get a strong first order phase transition.  Unlike
previous studies, we obtain the CP violation needed for baryogenesis
from renormalizable interactions of the scalar with hidden sector particles, 
in particular with the dark matter. In this
way we avoid the need for unspecified new physics at a low scale, and alleviate
CP constraints from searches for electric dipole moments. 
To our knowledge, this is the first example
of a model of electroweak baryogenesis where dark matter plays an essential
role in providing the initial CP asymmetry. We introduce the
notion of a ``CP portal'' interaction to transmit the CP asymmetry created in the
dark sector to standard model particles, as needed to bias sphalerons to
produce the baryon asymmetry. This requires new particle content (an inert 
Higgs doublet that can decay into dark matter and leptons)
which is near the discovery potential of LHC.  In the present model, the CP portal
coupling also determines the dark matter relic density.

In this preliminary study, we have computed the baryon asymmetry
quantitatively, while making a number of simplifying assumptions that
could be relaxed in future work.  The VEV of the scalar $S$ was taken
to vanish in the  true vacuum, but as noted above, there should be at
least a small VEV to  avoid domains containing cancelling
contributions to the baryon asymmetry; hence the $S\to -S$ symmetry in
the scalar potential should be broken.  It would be worthwhile to  to
quantify under what conditions the  higher-energy false vacua have
time to be diluted away before they disappear by tunneling to the
electroweak symmetry breaking vacuum. Breaking of the $S\to -S$
symmetry could also introduce explicit CP violation into the coupling
of $S$ to dark matter, as oppposed to the purely spontaneous violation
assumed here. Moreover it could lead to potentially observable signals for
direct dark matter detection as explained in section \ref{ddsec}, and
collider constraints from mixing of the Higgs with the singlet. 

Futher, we have ignored possible couplings of the inert  Higgs doublet $\phi$ to the SM Higgs, and
couplings of $\phi$ and $\chi$ to the $\mu$ and $e$ lepton doublets, which may be
constrained by lepton flavor  violating observables.  And it would be interesting
to determine and solve the renormalization group equations for the new couplings to
see at how high a scale the model is valid before encountering a Landau pole.

Regarding the phase transition, a number of improvements could be made. These
include dynamical solution of the bubble wall profiles in field space, which we
took to follow the minimum of the potential valley in the $(H,S)$ plane.
More importantly, we have assumed a value for the bubble wall velocity,
$v_w = 0.3\,c$, whereas an actual calculation would be desirable.  
Such determinations are notoriously difficult and hampered by
fundamental uncertainties in microscopic evaluation of the friction on the
wall~\cite{Moore:2000wx}; hence we leave this for future investigation.
Finally one could improve on the determination of the
elastic scattering rates, by including all  relevant channels. However, we already
addressed the most pressing issue about the rates in 
appendix~\ref{elastic_rates}, where we pointed out that the dominant contributions
are formally infrared-divergent, and must be regularized by a self-consistent
determination of the thermal scattering rate for a fermion of fixed energy in the
plasma.

Beyond technical improvements in the treatment of the model presented here, we
believe that the general framework, involving a singlet scalar and new states
connecting the scalar to standard model fermions, opens the door to building other
new models of electroweak baryogenesis, possibly involving dark matter, and
testable at the LHC. \\

{\bf Acknowledgments.}  We thank Joshua Berger, Simon Caron-Huot, Felix
Kahlhoefer, Guy Moore and Pat Scott for
helpful discussions. We thank Jonathan Cornell for assistance with
CosmoTransitions.  This work was financially supported by the Academy of Finland project 278722 
and by NSF grant 1216168.  Part of this work was performed at the Aspen Center for Physics, which 
is supported by NSF grant PHY-1066293.

%%%%%%%%%%%%%%%%%%%%%%%%%%%%%%%%%%%%%%%%%%%%%%%%%%%%%%%%%%%%%%%%%%%%%%%%%%%%%%%%%%%%%%%%%%%%%%%%%%%%
%%%%%%%%%%%%%%%%%%%%%%%%%%%%%%%%%%%%%%%%%%%%%%%%%%%%%%%%%%%%%%%%%%%%%%%%%%%%%%%%%%%%%%%%%%%%%%%%%%%%
%
\appendix
%
%%%%%%%%%%%%%%%%%%%%%%%%%%%%%%%%%%%%%%%%%%%%%%%%%%%%%%%%%%%%%%%%%%%%%%%%%%%%%%%%%%%%%%%%%%%%%%%%%%%%
%%%%%%%%%%%%%%%%%%%%%%%%%%%%%%%%%%%%%%%%%%%%%%%%%%%%%%%%%%%%%%%%%%%%%%%%%%%%%%%%%%%%%%%%%%%%%%%%%%%%

%---------------------------------------------------------------------------------------------------
%
\begin{figure}[t]
\hspace{-0.4cm}
\centerline{\includegraphics[width=0.85\hsize]{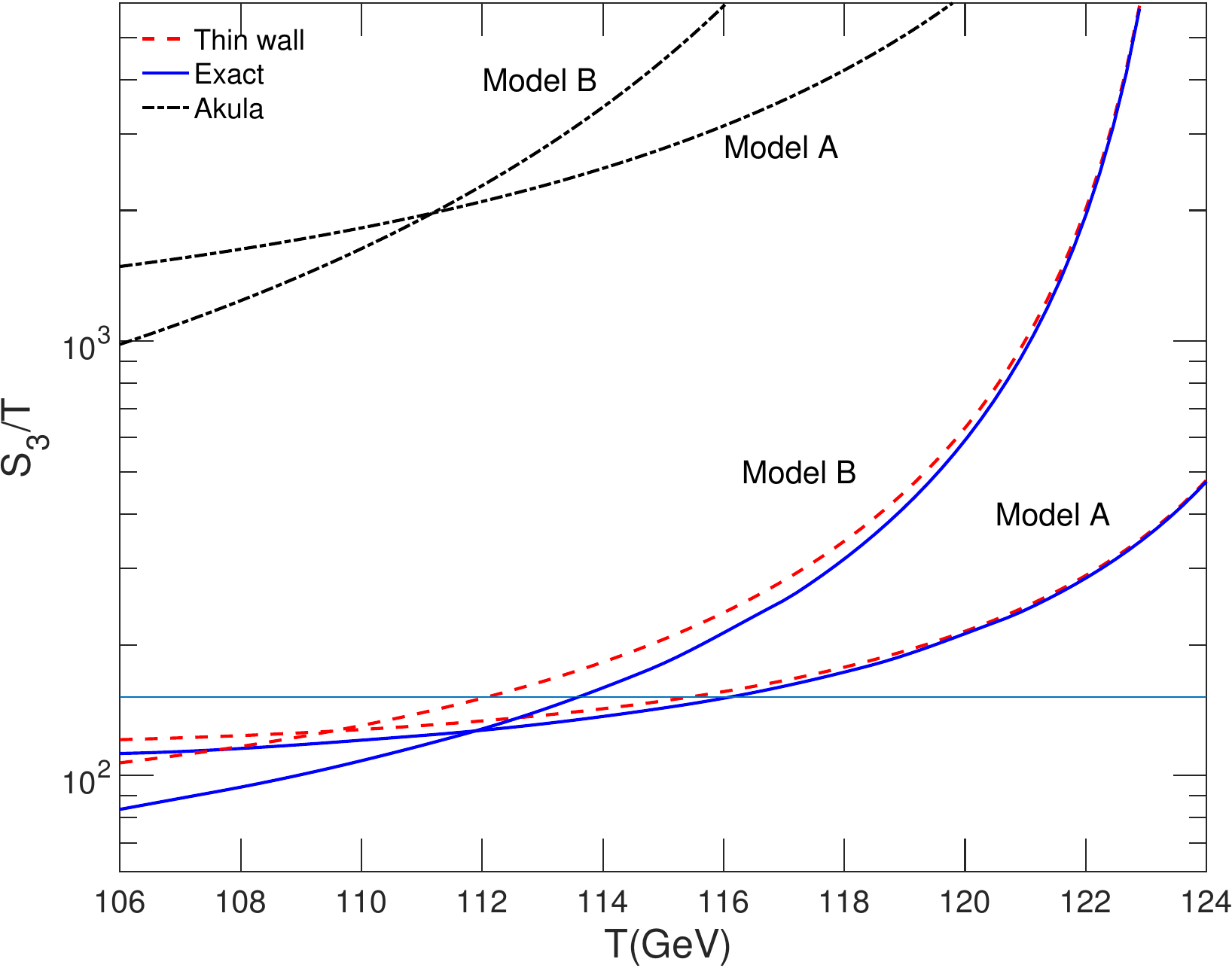}}
\caption{
Solutions for extremal action $S_3(T)/T$ for two different models as a
function of $T$. Model A (the benchmark model) represents typical
accuracy of our thin-wall approximation (red dashed lines), while
model B is representative of cases where the thin-wall action deviates
maximally from the exact action (blue solid lines). Black dash-dotted lines
give the leading order analytic approximation from
ref.~\cite{Akula:2016gpl}. Horizontal lines (overlapping) mark (exact)
$S_3(T_n)/T_n$ for both models.}
\label{fig:action}
\end{figure}
%
%---------------------------------------------------------------------------------------------------

\section{Improved thin-wall approximation for two fields}
\label{twapp}

We found the tunnelling solutions for the action (\ref{action}) by 
numerically solving the field equations
\be
\frac{d^2 \phi_i}{dr^2} + \frac{2}{r}\frac{d\phi_i}{dr} = \frac{dV}{d\phi_i} \,,
\label{eq:eom}
\ee
where $\phi_i=h,s$, with boundary conditions
$\frac{d\phi_i}{dr}|_{r=0}=0$ and $(h,s)\stackrel{{\scriptscriptstyle
r\rightarrow \infty}}{\longrightarrow} (0,w_T)$, where $(0,w_T)$ is
the $T$-dependent false minimum at high temperature. Numerical
solution is straightforward, but  takes several seconds per model,
whereby it is not ideally suited to extensive scans over model
parameters. Hence we initially tried using a (lowest order)
semianalytic formula of ref.~\cite{Akula:2016gpl} to approximate the
nucleation action.   Further investigation showed that this is a
poor approximation, typically overestimating action by a factor of 10
or more; see Fig.~\ref{fig:action}. 

We subsequently found that
$S_3/T$ can be well approximated by the following, slightly modified
version of the well known thin-wall
approximation~\cite{Coleman:1977py}. Assume that the tunneling path
follows a direct line between the two minima (this assumption is
supported by full numerical solutions; see Fig.~\ref{fieldpath}). 
This
reduces the problem to a single effective field
with potential $\tilde V(h,T) =
V(h,s(h),T)$, where $s\equiv w_T(1-h/v_T)$. The thin wall action then
takes the usual form:
\be
S_{\rm 3tw}(T) \approx \frac{16 \pi}{3}\frac{\sigma(T)^3}{\Delta V(T)^2} \,,
\label{eq:action}
\ee
where $\Delta V \equiv V(0,w_T,T)-V(v_T,0,T)$ and the surface tension is defined as as:
\be
\sigma(T) \equiv \sqrt{1 + \frac{w_T^2}{v_T^2}}\int_0^{v_T} dh \; 
{\rm Re} \left( \sqrt{2\delta V(h,T)} \, \right)\,,
\label{eq:sigma}
\ee
with $\delta V(h,T) \equiv \tilde V(h,T) - \tilde V(0,T) +
(h/v_T)\Delta V(T)$. The formulae (\ref{eq:action}-\ref{eq:sigma})
take essentially no time to numerically evaluate and provide surprisingly accurate
results, making this an ideal method for the initial model
selection in an extensive parameter scan, or more generally whenever a
very high precision is not needed. In Fig.~\ref{fig:action} we show a
comparison of the thin wall and exact numerical results for some
representative cases.  For comparison we also show the leading order
semianalytic result by ref.\ \cite{Akula:2016gpl} (black
dash-dotted curve).

%%%%%%%%%%%%%%%%%%%%%%%%%%%%%%%%%%%%%%%%%%%%%%%%%%%%%%%%%%%%%%%%%%%%%%%%%%%%%%%%%%%%%%%%%%%%%%%%%%%%%
%%%%%%%%%%%%%%%%%%%%%%%%%%%%%%%%%%%%%%%%%%%%%%%%%%%%%%%%%%%%%%%%%%%%%%%%%%%%%%%%%%%%%%%%%%%%%%%%%%%%
%
\section{Elastic scattering rates}
\label{elastic_rates}
%
%%%%%%%%%%%%%%%%%%%%%%%%%%%%%%%%%%%%%%%%%%%%%%%%%%%%%%%%%%%%%%%%%%%%%%%%%%%%%%%%%%%%%%%%%%%%%%%%%%%%
%%%%%%%%%%%%%%%%%%%%%%%%%%%%%%%%%%%%%%%%%%%%%%%%%%%%%%%%%%%%%%%%%%%%%%%%%%%%%%%%%%%%%%%%%%%%%%%%%%%%

Although there are many diagrams contributing to the scattering rates in fig.\ \ref{reactions}, 
they turn out to be dominated by just a few. These are the diagrams with an external $\phi$, 
where an internal $\tau$ or $\chi$ particle in the $t$ or $u$ channel can go on shell and ones
with external $\tau$ and $\chi$ where $\phi$ can go on-shell in $s$-channel. 
In addition the $t$ and $u$-channel diagrams contributions are IR divergent. All these
diagrams must be regularized by taking into account the thermal damping rate $\Gamma_x$ of the
intermediate particle $x$. It amounts to using the propagator \cite{Notzold:1987ik}
\be
 {i\over \slashed{p} - m_x +  i \Gamma_x \slashed{u}/2} \,,
\label{therm_prop}
\ee
where $u$ is the plasma 4-velocity. (This is the leading correction to the imaginary part of the
self-energy from thermal corrections.)  The rationalized propagator then takes the form
\be
i{\slashed{p} + m_x \over p^2-m_x^2 + i\epsilon}
\label{thermal_prop2}
\ee
where $\epsilon = p_0 \Gamma_x$.  At lowest order the thermal damping rates are given by the
decay rates computed in appendix~\ref{decays}: $\Gamma_i = I_d/n_i$.

Moreover, we need to subtract the on-shell contribution to avoid double-counting the decays and 
inverse decays of $\phi$, which have already been explicitly included in the Boltzmann equations.  
The need for such subtractions has been discussed in similar contexts 
\cite{Kolb:1979qa,Cline:1993bd,Giudice:2003jh}.  However there is some
ambiguity in the literature as to how this should be done.   A widely
used prescription is to replace the squared propagator $|P(p)|^2$ with
\be
  |P(p)|^2 \to  {1\over (p^2-m^2)^2 + m^2\Gamma^2} - {\pi\over m\Gamma} \delta(p^2 - m^2)
\label{subtraction0}
\ee
in the case of a massive intermediate particle that has a decay width $\Gamma$.
It has been noted \cite{Cline:1993bd} that this can lead to negative cross
sections.  Ref.\ \cite{Giudice:2003jh} proposes a different method of
subtraction that avoids this problem, but it is specific to $s$-channel processes and does
not suffice for us.

%---------------------------------------------------------------------------------------------------
%
\begin{figure}[t]
\centerline{
\hskip-0.4cm \includegraphics[width=0.7\hsize]{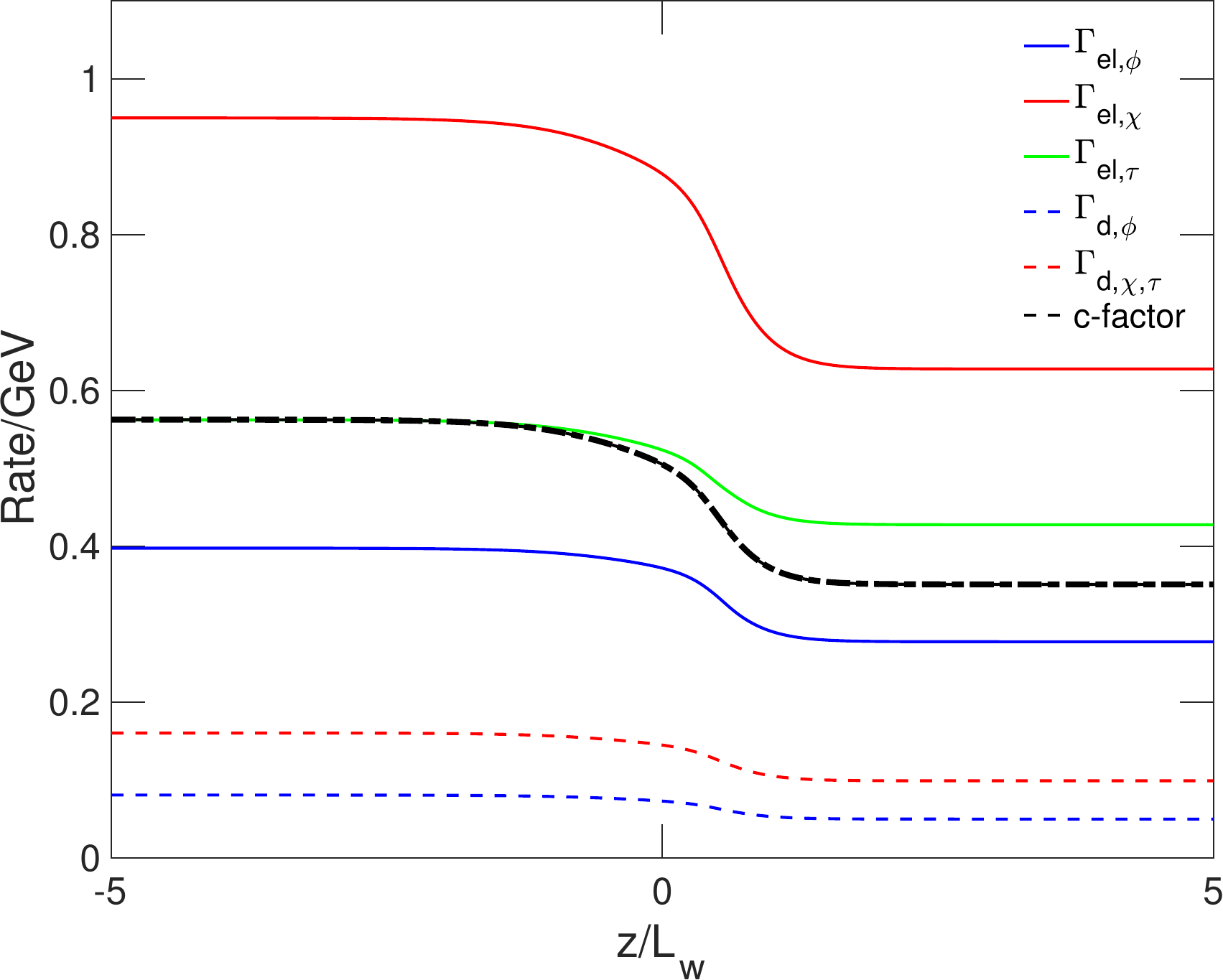}}
\caption{Total elastic rates (solid lines) and decay rates (dashed lines) 
as a function of distance from bubble wall $z/L_w$ 
for our benchmark model.
The decay helicity-flip $c$-factor discussed in appendix \ref{decays}
(eq.\ (\ref{c-factor})) is given by the thick dash-dotted line.}
\label{fig:elasticrates}
\end{figure}
%
%---------------------------------------------------------------------------------------------------

We propose to carry out a subtraction similar to (\ref{subtraction0}), but rather than doing 
so on the squared propagator appearing in the cross section, we apply it to the propagator in 
the amplitude. By construction this avoids the problem of a negative cross 
section and it also seems to be better justified physically. Indeed the idea is that the pole 
contribution gets associated with a distinct physical process: a successive decay and inverse decay
of a particle, which is not part of the $2\leftrightarrow 2$ scattering matrix 
but instead of the $1\leftrightarrow 2$ matrix. It should then be removed completely, 
including all interference effects, which can only be ensured by subtracting at the amplitude 
level.  For a scalar particle, our prescription is to take the
principal value part 
of the propagator by removing the real part
\be
	P(p) \to  i{p^2 - m^2 \over (p^2-m^2)^2 + \epsilon^2}
\label{subtraction1}
\ee
where $\epsilon = m\Gamma$ for a particle with a decay width. For a fermion, we multiply 
(\ref{subtraction1}) by $\slashed{p} + m$.

%---------------------------------------------------------------------------------------------------
%
\begin{figure*}[t]
\hspace{-0.4cm}
\centerline{ \includegraphics[width=0.4\hsize]{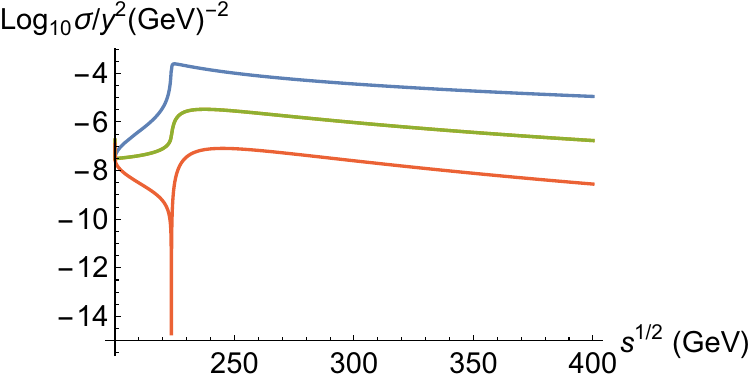}
             \includegraphics[width=0.4\hsize]{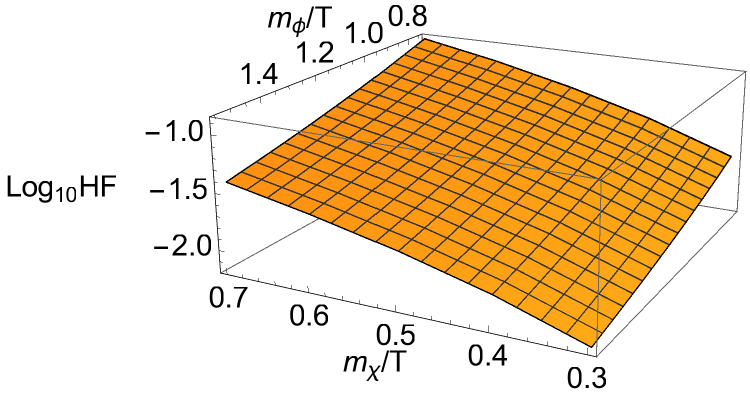}}
\caption{Left: cross sections for helicity-conserving $\lambda_i=+1$
(top), helicity-flipping (middle) and helicity-conserving
$\lambda_i=-1$ $\chi\phi$ scattering, versus $\sqrt{s}$, for 
$m_\chi=50\,$GeV, $m_\phi = 150\,$GeV, and $y=0.75$ in the thermal
width of the intermediate $\tau$.  Right: ratio of $\lambda_1 =
-\lambda_4 = \pm 1$ helicity-flipping cross sections to that of the
dominant $\lambda_i=+1$ helicity-conserving one, as a function of 
$m_\phi/T$ and $m_\chi/T$.  It is nearly independent of $y$.}
\label{hflip}
\end{figure*}
%
%---------------------------------------------------------------------------------------------------

The dominant contribution to the cross section for $\chi\phi$ ($\tau\phi$) scattering, 
due to $t$-channel exchange, is controlled by the thermal width of $\tau$
($\chi$).
Similarly the cross-section for $\chi\tau$ is dominated by the $s$-channel diagram controlled
by thermal width of $\phi$. Contributions from these channels are, even after pole 
subtractions, as large or larger than the decay contribution. That is, we are finding that
the two-loop contributions to the complex part of the self-energy are very large, which means 
that we must solve a gap equation for the thermal widths involving all three particles and 
their dominant decay and scattering channels.
These equations can be formally written as
\be
\Gamma_i = \Gamma_{d,i} + \Delta_i[\Gamma_j]
\label{eq:gap}
\ee
where the $\Delta_i[\Gamma_j]$ arise from thermally averaged scattering rates:
\be
\Delta_{i}[\Gamma_j] = \sum_k n_k \langle v_{\rm rel}
\sigma_{ik}[\Gamma_j]\rangle
\ee
As discussed above, we include here the dominant processes: $\chi\phi \rightarrow \chi\phi$,
$\phi\tau \rightarrow \phi\tau$ and $\chi\tau \rightarrow \chi\tau$. In addition we included 
$f\tau \rightarrow f\tau$ and $\bar f\tau \rightarrow \bar f\tau$,
where $f$ denotes standard model fermions. (Although these are
IR convergent, taking account of the thermal mass of the $W$ 
$m^2_W = (8/3)g^2 T^2$ in 
$t$-channel exchange, there are many flavors.)   When the dominant 
scattering channels are included along with the decays, the infrared sensitivity of independent 
channels is greatly reduced, and results stabilize against adding new contributions. For example
adding $f\tau$ and $\bar f\tau$ channels does not change the results dramatically despite their
large multiplicity. 

The calculation of scattering rates is straightforward and we do note
show the lengthy expressions for the cross-sections (computed using
FeynCalc  \cite{Shtabovenko:2016sxi}) here.  We determined  all thermal averages using
\be
 \langle v_{\rm rel}\sigma_{ik}\rangle = \int_{\sss(m_i+m_k)^2}^\infty \hskip-0.4cm
 {\sqrt{s}\lambda^{1/2}(s,m_i^2,m_k^2)
 \,K_1(\frac{\sqrt{s}}{T})\, v_{\rm rel}\sigma_{ik}  \over 16 Tm_i^2m_k^2\,
 K_2(\frac{m_i}{T})K_2(\frac{m_k}{T})}\,{\rm d}s.
\label{thermsv}
\ee
The behaviour of the solutions of the equation (\ref{eq:gap}) in the
bububble wall, along with the decay rates
$\Gamma_{d,i}$, is shown  in figure (\ref{fig:elasticrates}) for our 
benchmark model. Note that diffusion
properties change significantly from one phase to another.

There are several other scattering channels that have IR enhancement, but are still 
subdominant. These are $\chi\chi\to S^*\to\chi\chi$ annihilation, if $m_\chi < m_S/2$ with 
the possibility of an $s$-channel resonance; similarly $L_\tau\bar L_\tau\to W^*  \to 
L_\tau\bar L_\tau$, since the thermal mass of the $W$ in the symmetric phase is more than 
two times greater than that of  $L_\tau$; and $L_\tau W\to L_\tau W$, where the $W$ decays 
to $L_\tau\bar L_\tau$ followed by inverse decay, similar to  $L_\tau \phi\to L_\tau
\phi$ discussed above.  The first process tends to be absent in our
scans, which favor $m_\chi > m_S/2$ for getting the observed BAU and
DM density, and is also suppressed by $\eta/y\ll 1$ relative to the
dominant processes.  The others  which are proportional to $g^2$ turn
out to be numerically small compared to the dominant $y^2$ processes.

%%%%%%%%%%%%%%%%%%%%%%%%%%%%%%%%%%%%%%%%%%%%%%%%%%%%%%%%%%%%%%%%%%%%%%%%%%%%%%%%%%%%%%%%%%%%%%%%%%%%
%%%%%%%%%%%%%%%%%%%%%%%%%%%%%%%%%%%%%%%%%%%%%%%%%%%%%%%%%%%%%%%%%%%%%%%%%%%%%%%%%%%%%%%%%%%%%%%%%%%%
%
\section{Helicity flip rate}
\label{hf_rates}
%
%%%%%%%%%%%%%%%%%%%%%%%%%%%%%%%%%%%%%%%%%%%%%%%%%%%%%%%%%%%%%%%%%%%%%%%%%%%%%%%%%%%%%%%%%%%%%%%%%%%%
%%%%%%%%%%%%%%%%%%%%%%%%%%%%%%%%%%%%%%%%%%%%%%%%%%%%%%%%%%%%%%%%%%%%%%%%%%%%%%%%%%%%%%%%%%%%%%%%%%%%

The rate of helicity flips in scattering can be computed by insertion of the helicity
projection operators 
\vskip-0.3cm
\be
	\Sigma_i = \sfrac12 (1 + \gamma_5\slashed{s}_i)
\ee
into the amplitudes involving $\chi$ scattering, 
where the spin vector of the $i$th particle is $s_i^\mu = \lambda_i (|p_i|,
E_i \hat p_i)$ with $\lambda_i = \pm 1$ for positive or negative helicity.
Carrying this out for $\chi\phi\to\chi\phi$ scattering, we find that
the largest rate is for the helicity-conserving case where 
$\lambda_1 = \lambda_4 = +1$ (labeling the incoming and outgoing
$\chi$ particles by $i=1,4$ respectively), 
followed by the two helicity-flipping possibilities (which have equal
cross sections), and then
$\lambda_1 = \lambda_4 = -1$.  The distinction between the first and
last is due to scattering on $\phi$ as opposed to $\phi^*$.  
We show the dependence of the cross sections on energy in fig.\
\ref{hflip} (left).  

Carrying out the thermal averages, the ratio of the rates for either
helicity flip process versus the dominant helicity-conserving one 
is a function of $m_\chi/T$ and $m_\phi/T$ which to a good
approximation is independent of $y$, and is plotted in fig.\
\ref{hflip} (right).  For numerical purposes, we find that it can
be accurately approximated (to better than 1\% for the parameters
of interest) by a second order polynomial in $m_\chi$ and $m_\phi$,
whose coefficients depend very weakly on $y$.  
For our benchmark model, the total flip ratio is 8\%.

%%%%%%%%%%%%%%%%%%%%%%%%%%%%%%%%%%%%%%%%%%%%%%%%%%%%%%%%%%%%%%%%%%%%%%%%%%%%%%%%%%%%%%%%%%%%%%%%%%%%
%%%%%%%%%%%%%%%%%%%%%%%%%%%%%%%%%%%%%%%%%%%%%%%%%%%%%%%%%%%%%%%%%%%%%%%%%%%%%%%%%%%%%%%%%%%%%%%%%%%%
%
\section{Decay rates}
\label{decays}
%
%%%%%%%%%%%%%%%%%%%%%%%%%%%%%%%%%%%%%%%%%%%%%%%%%%%%%%%%%%%%%%%%%%%%%%%%%%%%%%%%%%%%%%%%%%%%%%%%%%%%
%%%%%%%%%%%%%%%%%%%%%%%%%%%%%%%%%%%%%%%%%%%%%%%%%%%%%%%%%%%%%%%%%%%%%%%%%%%%%%%%%%%%%%%%%%%%%%%%%%%%

\begin{figure*}[t]
\hspace{-0.4cm}
\centerline{
{\includegraphics[width=0.37\hsize]{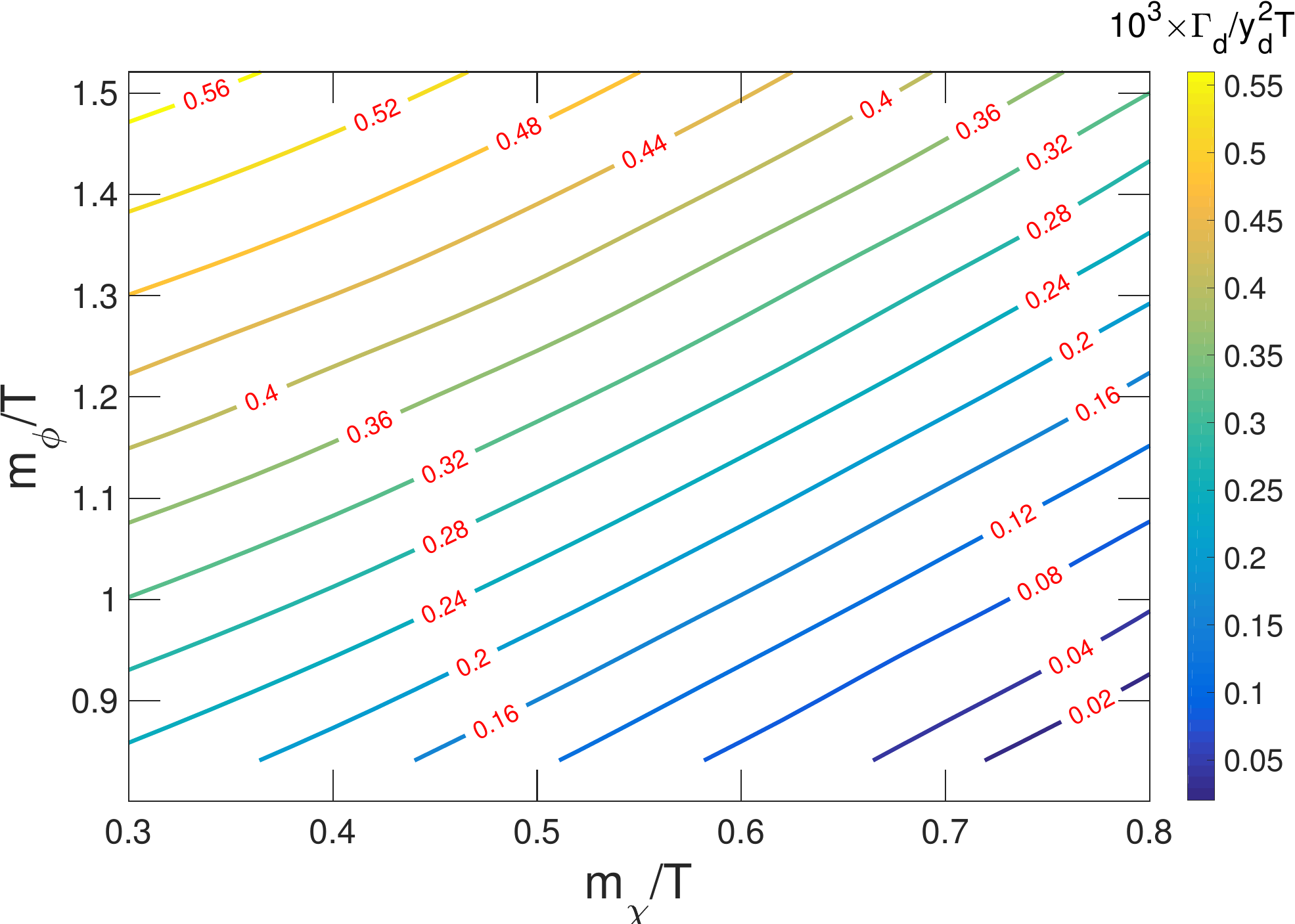}}\hfil
 \includegraphics[width=0.36\hsize]{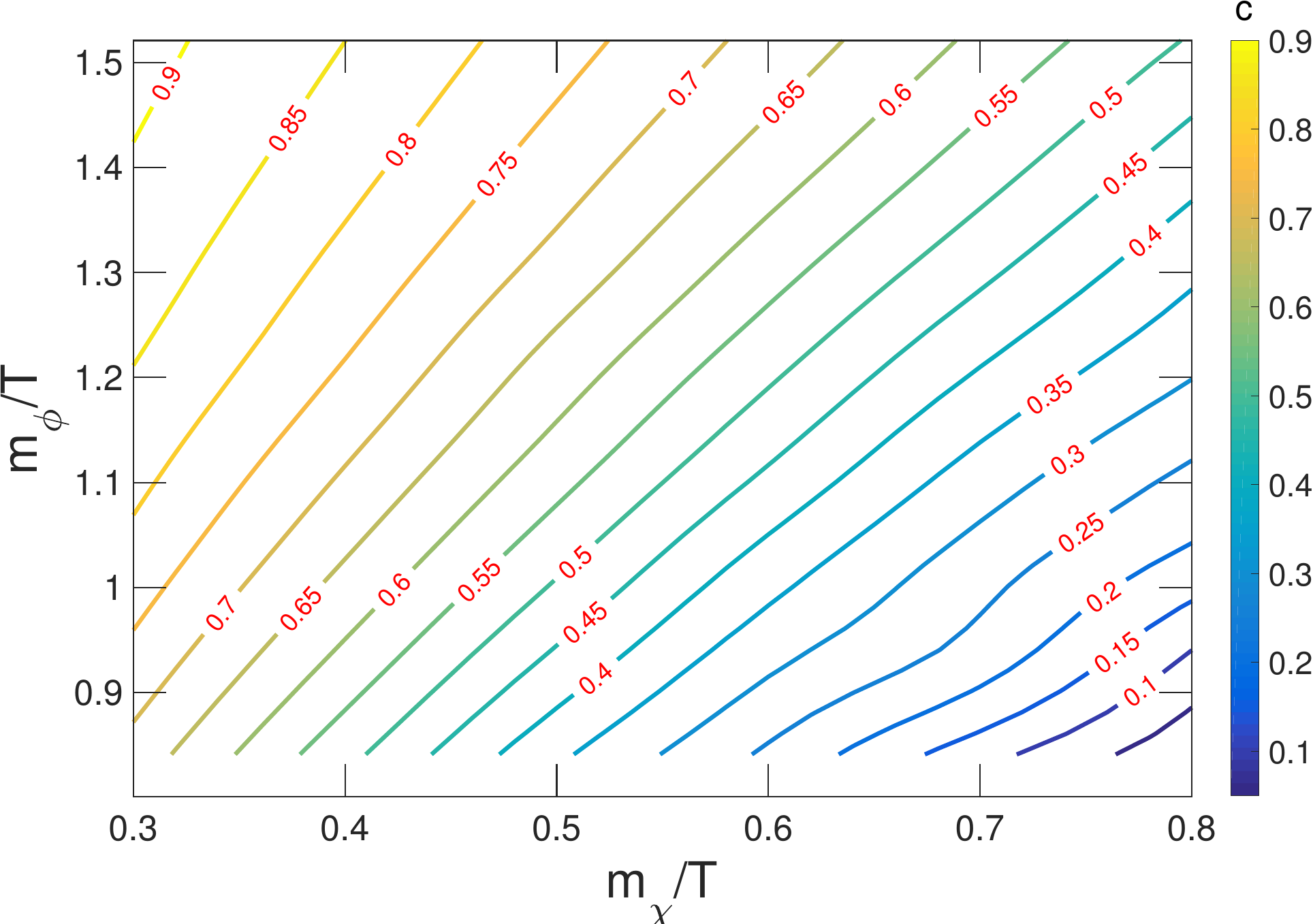}}
\caption{Left: contours of the thermally averaged, suitably scaled $\phi$ decay rate $10^3 \Gamma_d/y^2 T$ as a function of 
$m_\chi/T$ and $m_\phi/T$. Right: contours of the $\chi$ 
helicity-flip factor $c$, eq.\ (\ref{c-factor}).}
\label{decay}
\end{figure*}

Here we derive thermally averaged decay rates for 
 $\phi_\pm \rightarrow \chi_h \tau_\pm$, with $h$ denoting the
helicity of the $\chi$.  
It is straightforward to show that the collision integral in the Boltzmann
equations, when expanded to first order in the
chemical potentials, becomes
\be
C_\pm^h(\mu) = \left(  \mp \frac{\mu_\phi}{T} \pm \frac{\mu_\tau}{T} - h \frac{\mu_\chi}{T}\right)
\times I_{h\pm} + ...\,,
\label{eq:chpm}
\ee
where dots represent terms that contribute to elastic integrals and
\bea
I_{h\pm} &=& 2y^2\int_p\int_k\int_q (2\pi )^4\delta^4(p-k-q)\times
\nn \\&& \times \Big(k\cdot q \pm M_\chi s_h \cdot k \Big)
 (1-f_{0,\phi}(p)) f_{0,\chi}(q)f_{0,\tau}(k)
\nn \\
&\equiv& I_d \pm h I_h\,.
\label{eq:Is}
\eea
Here $\int_{p_i} \equiv \int {\rm d}^3p_i/[(2\pi)^32\omega_i]$, 
$f_{0,i}(p_i)$ are the Bose-Einstein or Fermi-Dirac
equilibrium distributions for $\phi$, $\chi$ and $\tau$, and the 
spin 4-vector has components 
$s_h = (h/m_\chi)(|{\bf q}|,\omega_q\hat{\bf q})$.  Using Lorentz 
invariance of $k\cdot q$ we can reduce 
first term $I_d$ in eq.\ (\ref{eq:Is}) to a 1D-integral,
\bea
&&	I_d = {y^2(m_\phi^2-m_\chi^2)m_\phi T\over 32\pi^3}\int_1^\infty du
	{e^{m_\phi u/T}\over (e^{m_\phi u/T}-1)^2} \cdot\\ 
&\cdot&\ln\left[
	\cosh(a_+u+a_-\sqrt{u^2-1}) \cosh(a_-(u+ \sqrt{u^2-1}))\over
	\cosh(a_+u-a_-\sqrt{u^2-1})\cosh(a_-(u- \sqrt{u^2-1}))\right],\nn
\eea
where $a_\pm = (m_\phi^2\pm m_\chi^2)/(4 m_\phi T)$. It is easy to see that $\Gamma_d \equiv 
I_d/n_\phi$ is the $\phi$ decay rate, which reduces to the vacuum rate in the $T\to 0$ limit.
$\Gamma_d$ can be expressed as $y^2 T$ times a function of $m_\chi/T,m_\phi/T$, which is
shown in fig.~\ref{decay} (left).  Because $s_h \cdot k$ is not 
Lorentz invariant, the second term $I_h$ can 
only be reduced to a two-dimensional integral in the most general 
case and we do not reproduce the result
here.

%---------------------------------------------------------------------------------------------------
%
%
%---------------------------------------------------------------------------------------------------

These decay collision terms give rise to the following 
contributions to $C^\mu_i$-terms appearing 
in the fluid equations (\ref{diffeqs}) (superscript $\mu$ denotes
chemical potential):
\bea
\delta C^\mu_j &=& \frac{1}{2N_j}\sum_h(C^h_+-C^h_-), \quad j=\phi,\tau \nn\\
\delta C^\mu_\chi &=& \frac{g_w}{2N_\chi}\sum_\pm(C^+_\pm-C^-_\pm)
\label{eq:secondCeq}
\eea
where $N_i$ are the zero-mass normalization factors defined for $\chi$ below equation 
(\ref{eq:Knfuns}). The coefficient $g_w=2$ in $\delta C^\mu_\chi$ is the SU(2)-multiplicity factor.

Using equations (\ref{eq:chpm}) and (\ref{eq:Is}), we can write 
(\ref{eq:secondCeq}) as
\bea
\delta C^\mu_\phi &=& \frac{I_d}{TN_\phi} (\mu_\phi - \mu_\tau - c\mu_\chi ) \nn\\
\delta C^\mu_\chi &=& \frac{I_d}{TN_\chi} (\mu_\chi + c\mu_\tau + c\mu_\phi ) \nn\\
\delta C^\mu_\tau &=& \frac{2I_d}{TN_\tau}(\mu_\tau - \mu_\phi + c\mu_\chi ) \,,
\eea
where the helicity-flipping factor is defined as
\be
c \equiv I_h/I_d \,.
\label{c-factor}
\ee
We then infer the effective decay rates appearing in (\ref{diffeqs}), 
\be
\tGamma_\phi = \frac{I_d}{TN_\phi}= \frac{n_\phi}{TN_\phi}\Gamma_\phi \equiv r_\phi\Gamma_\phi
\label{eq:gtilde}
\ee
and similarly for $\chi$ and $\tau$. The normalization factors $r_i$ are induced here (and 
in all other rates in (\ref{diffeqs})) by the particular normalization chosen for the 
$K_{i,j}$ functions.

The decay processes are crucial to our CP-portal mechanism as they are the only way to bring 
the $\chi$ CP-asymmetry into the $\tau$ sector. In the limit $m_\chi \rightarrow 0$ the $\chi$-helicity 
follows faithfully the chirality of $\tau$ and the transport is maximally efficient; this
corresponds to the case $c=1$. However, with a nonzero $m_\chi$ some decays produce wrong 
helicity $\chi^\prime$s, whereby $c<1$. For small $c$ the transport is less efficient and in the limit 
$c\rightarrow 0$, $\mu_\chi$ completely decouples, hiding CP-violation in the dark sector. 
We show $c$ as a function of $m_\chi/T,m_\phi/T$ in fig.~\ref{decay} (right): it turns out that
$c$ is strongly suppressed only in the limit that $\phi$ and  $\chi$-masses become 
degenerate, {\em i.e.}~in the limit that $\chi$ is nonrelativistic (as expected).

%%%%%%%%%%%%%%%%%%%%%%%%%%%%%%%%%%%%%%%%%%%%%%%%%%%%%%%%%%%%%%%%%%%%%%%%%%%%%%%%%%%%%%%%%%%%%%%%%%%
%
\section{Semiclassical source in helicity basis}
\label{sec:source}
%
%%%%%%%%%%%%%%%%%%%%%%%%%%%%%%%%%%%%%%%%%%%%%%%%%%%%%%%%%%%%%%%%%%%%%%%%%%%%%%%%%%%%%%%%%%%%%%%%%%%%

In this section we derive the semiclassical source in the helicity
basis. Here we label states by their wall-frame helicities, whereas we
use plasma-frame helicities in collision-term calculations.  This is
justified given that we calculate the BAU to first-order in $v_w$.  
 We start from the 3D 
semiclassical force derived in~\cite{Kainulainen:2002th}:
\be
F^s_{\chi} = - \frac{{m_\chi^2}^{\,\prime} }{2k_0 }
             + s_{\sss CP} \frac{s(m_\chi^2\theta^{\,\prime})^{\,\prime}}{2k_0 k_{0\sparallel}},
\label{Fspm}
\ee
where $k_{0\sparallel} = ({k_0^2-{\bf k}_\sparallel^{\,2}})^\frac 12$
and $s_{\sss CP} = 1\,(-1)$ for 
particles (antiparticles) and $s$ refers to the spin perpendicular to the wall in the frame where
the particle velocity parallel to wall vanishes, ${\bf k}_\sparallel \equiv 0$. That is, (\ref{Fspm})
is the force acting on the ${\bf k}_\sparallel \equiv 0$-frame eigenstate of the spin operator, 
$S_z = \gamma^0\gamma^3\gamma^5$, boosted back to wall frame, or equivalently, it is the
force acting on the eigenstates $u(p,s)$ of the boosted spin-operator~\cite{Kainulainen:2002th}
\begin{equation}
  S^{\rm wf}_z  = \gamma_\sparallel \Big( S_z 
      - i({\bf v}_\sparallel \times \alpha)_z \Big) \,.
\label{Sz}
\end{equation}
where superscript ``wf'' indicates the rest frame of the bubble wall.
To get the force on helicity states, we first note that we can always write a helicity state as a 
linear combination of $u(p,s)$ spin eigenstates: $u(p,h) = \sum_s c_s u(p,s)$. Note that force has opposite 
signs on opposite spins, which reflects the fact that at the quantum level $u(p,s=\pm 1)$ evolve 
under slightly different effective Hamiltonians. This leads to a separation of the wave 
function, analogously to the separation of electron spin states in the Stern-Gerlach experiment. 
However, our kinetic equations are classical, which by (molecular chaos) assumption 
forbid any correlations between initial states in collision events. 
This means that, consistently 
with the semiclassical picture, we must treat spin as a classical variable. 

\begin{figure}[t]
\hspace{-0.4cm}
\centerline{\includegraphics[width=0.85\hsize]{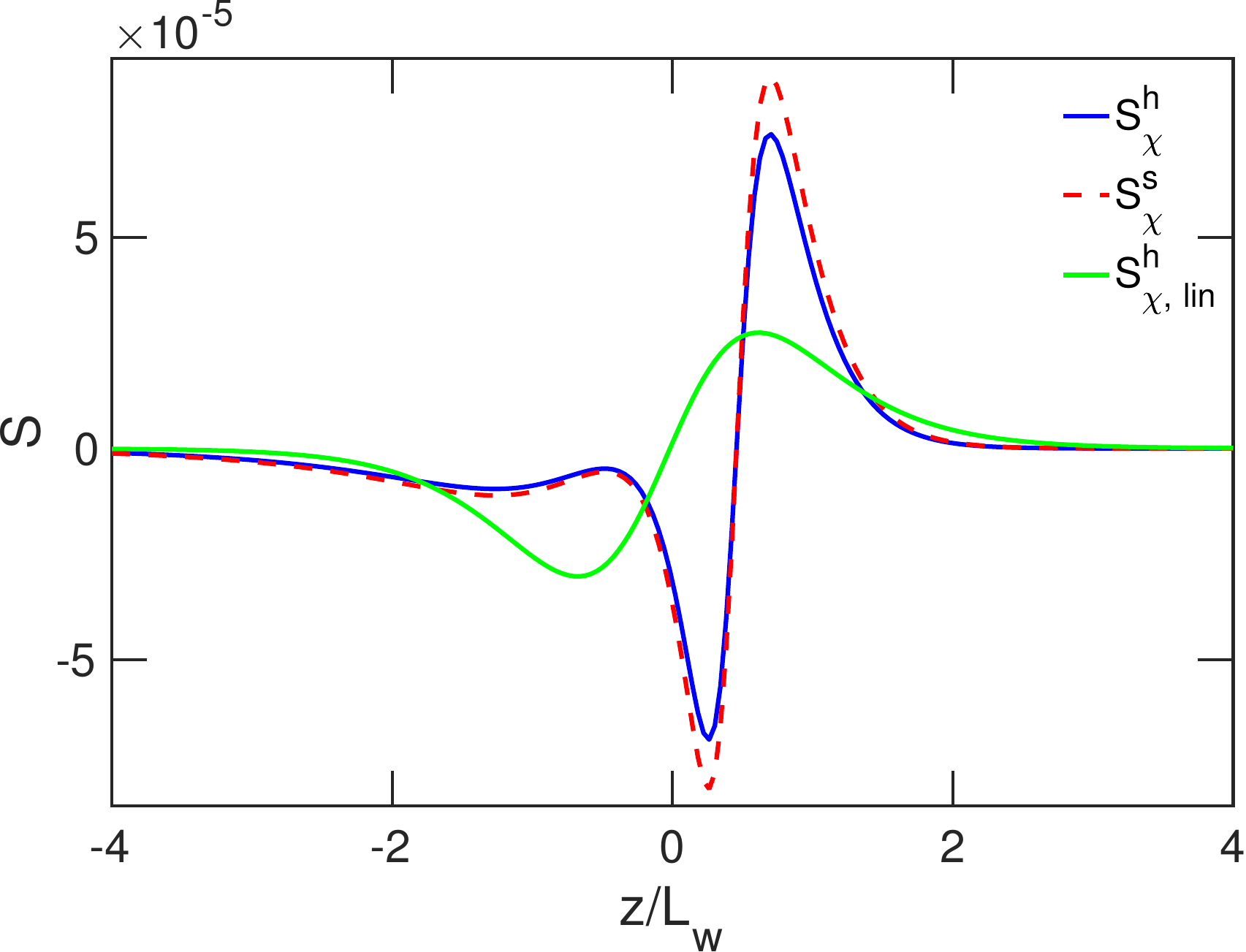}}
\caption{The helicity source function $S_\chi^h$ used in the
present work versus the spin 
basis source $S_\chi^s$ of ref.\ \cite{Fromme:2006wx} (highest two
curves).  The lowest curve is $S_\chi^h$ computed with a straight-line
approximation in $h$-$s$ field space to the bubble wall path, rather
than the more accurate curved path. }
\label{fig:sources}
\end{figure}

Thus the force 
acting on helicity states $u(p,h)$ is the quantum average of the force (\ref{Fspm}) over spin $s$. 
In practice this means replacing $s \rightarrow \langle s\rangle_h$ in (\ref{Fspm}) 
where $\langle s\rangle_h$ is the expectation value of the spin in the helicity
eigenstate:
\be
\langle s\rangle_h = \sum_s s|c_s|^2 
    = \langle p,h | S^{\rm wf}_z | p,h \rangle
    = \gamma_\sparallel \frac{h|k_z|}{|{\bf k}|} \,,
\ee
and the boost factor is $\gamma_\sparallel = k_0/k_{0\sparallel}$.
 We can now compute the source for 
the fluid equations (\ref{diffeqs}) arising from the force $F^h_\chi= F^{s=\langle s\rangle_h}_\chi$. 
Here we do not 
present the full calculation, performed in~\cite{Fromme:2006wx},
but rather just  emphasize one 
issue: in~\cite{Fromme:2006wx} thermal averages are {\em defined} by integrating a quantity $X$ 
over effective $k_z$-momentum, but with $k_0$ on shell, {\em e.g.},
\bea
\langle X \rangle_\pm &\equiv& \frac{1}{N_\chi}\int {\rm
d}^4k\,X(k_0;k_i)\,\theta(\pm k_0)
\nn \\
&& \phantom{Ha} \times |2k_0| \,
\delta\left(k_0^2-\omega_0^2 +
\langle s\rangle_h\frac{m_\chi^2\theta^\prime}{k_{0\sparallel}}\right) \nn \\
&\approx& \frac{1}{N_\chi} \int {\rm d}^3k\, 
X\left(\omega_0 \mp 
\langle s\rangle_h\frac{m_\chi^2\theta^\prime}{2\omega_0\omega_{0\sparallel}};k_i\right)
\nn \\
&\approx& \langle X(\omega_0;k_i) \rangle \mp 
\langle s\rangle_h m_\chi^2\theta^\prime\left\langle 
\frac{X^\prime(\omega_0;k_i)}{2\omega_0\omega_{0\sparallel}}\right\rangle\,,
\eea
where $N_\chi$ is a normalization factor, $X^\prime \equiv {\rm d}X/{\rm d}\omega_0$ and 
$\omega^2_0 = {\bf k}^2 + m_\chi^2$. 
This is the reason why also the CP-even part of the force~(\ref{Fspm}) gives rise to a CP-odd 
source (the $K^h_9$-source below). Working to leading order in gradients, one eventually finds the 
result (\ref{source}) in the main text, where 
\bea
K^h_8(x) &=& \left\langle
\frac{k_z^2f^\prime_0}{2\omega_{0}\omega_{0\sparallel}^2|{\bf
k}|}\right\rangle 
\nn\\
K^h_9(x) &=& \left\langle \frac{k_z^2}{4\omega_{0}^2\omega_{0\sparallel}^2|{\bf k}|}
\left(\frac{f_0^\prime}{\omega_0}-f_0^{\prime\prime}\right)\right\rangle \,,
\label{eq:Knfuns}
\eea
with $f^\prime_0 = {\rm d}f_0(\omega_0)/{\rm d}\omega_0$ and  $N_\chi
\equiv \int {\rm d}^3k f_0^\prime(m=0)$. The functions $K^h_{8,9}$ differ
from  those derived from the $S^s_\chi$-source, given
in~\cite{Fromme:2006wx}, but  in practice the difference between the
helicity-basis and spin-basis sources  is small. The effect is
shown in fig.~\ref{fig:sources} for our benchmark case. For
comparison, we show the change in the helicity source due to adopting
a straight line approximation between the $T_n$-minima in the field
space. 

%---------------------------------------------------------------------------------------------------
%
%
%---------------------------------------------------------------------------------------------------

%%%%%%%%%%%%%%%%%%%%%%%%%%%%%%%%%%%%%%%%%%%%%%%%%%%%%%%%%%%%%%%%%%%%%%%%%%%%%%%%%%%%%%%%%%%%%%%%%%%%
%
\bibliographystyle{apsrev}
%\bibliography{ref_abelian_dm,scalarDM,seesaw,ref4thGen}

\begin{thebibliography}{10}
%
%%%%%%%%%%%%%%% MSSM %%%%%%%%%%%%%%%%%%%%%%%%%%%%%%%%%%%%%%%%%%%%%%%%%%%%%%%%%%%%%%%%%%%%%%%%%%%%%%%

%\cite{Liebler:2015ddv,Carena:2012np,Curtin:2012aa}
\bibitem{Liebler:2015ddv} 
  S.~Liebler, S.~Profumo and T.~Stefaniak,
  ``Light Stop Mass Limits from Higgs Rate Measurements in the MSSM: Is MSSM Electroweak Baryogenesis Still Alive After All?,''
  JHEP {\bf 1604}, 143 (2016)
  doi:10.1007/JHEP04(2016)143
  [arXiv:1512.09172 [hep-ph]].
  %%CITATION = doi:10.1007/JHEP04(2016)143;%%
  %1 citations counted in INSPIRE as of 22 Aug 2016

%\cite{Carena:2012np}
\bibitem{Carena:2012np} 
  M.~Carena, G.~Nardini, M.~Quiros and C.~E.~M.~Wagner,
  ``MSSM Electroweak Baryogenesis and LHC Data,''
  JHEP {\bf 1302}, 001 (2013)
  doi:10.1007/JHEP02(2013)001
  [arXiv:1207.6330 [hep-ph]].
  %%CITATION = doi:10.1007/JHEP02(2013)001;%%
  %69 citations counted in INSPIRE as of 22 Aug 2016

%\cite{Curtin:2012aa}
\bibitem{Curtin:2012aa} 
  D.~Curtin, P.~Jaiswal and P.~Meade,
  ``Excluding Electroweak Baryogenesis in the MSSM,''
  JHEP {\bf 1208}, 005 (2012)
  doi:10.1007/JHEP08(2012)005
  [arXiv:1203.2932 [hep-ph]].
  %%CITATION = doi:10.1007/JHEP08(2012)005;%%
  %68 citations counted in INSPIRE as of 22 Aug 2016

%%%%%%%%%%%%%%%%%%%%%%% 2HDM %%%%%%%%%%%%%%%%%%%%%%%%%%%%%%%%%%%%%%%%%%%%%%%%%

%\cite{Cline:2011mm}
\bibitem{Cline:2011mm} 
  J.~M.~Cline, K.~Kainulainen and M.~Trott,
  ``Electroweak Baryogenesis in Two Higgs Doublet Models and B meson anomalies,''
  JHEP {\bf 1111}, 089 (2011)
  doi:10.1007/JHEP11(2011)089
  [arXiv:1107.3559 [hep-ph]].
  %%CITATION = doi:10.1007/JHEP11(2011)089;%%
  %55 citations counted in INSPIRE as of 22 Aug 2016

%\cite{Haarr:2016qzq}
\bibitem{Haarr:2016qzq} 
  A.~Haarr, A.~Kvellestad and T.~C.~Petersen,
  ``Disfavouring Electroweak Baryogenesis and a hidden Higgs in a CP-violating Two-Higgs-Doublet Model,''
  arXiv:1611.05757 [hep-ph].
  %%CITATION = ARXIV:1611.05757;%%

%\cite{Chiang:2016vgf}
\bibitem{Chiang:2016vgf} 
  C.~W.~Chiang, K.~Fuyuto and E.~Senaha,
  ``Electroweak Baryogenesis with Lepton Flavor Violation,''
  Phys.\ Lett.\ B {\bf 762}, 315 (2016)
  doi:10.1016/j.physletb.2016.09.052
  [arXiv:1607.07316 [hep-ph]].
  %%CITATION = doi:10.1016/j.physletb.2016.09.052;%%
  %3 citations counted in INSPIRE as of 03 Dec 2016

%\cite{Dorsch:2016nrg}
\bibitem{Dorsch:2016nrg} 
  G.~C.~Dorsch, S.~J.~Huber, T.~Konstandin and J.~M.~No,
  ``A Second Higgs Doublet in the Early Universe: Baryogenesis and Gravitational Waves,''
  arXiv:1611.05874 [hep-ph].
  %%CITATION = ARXIV:1611.05874;%%


%%%%%%%%%%%%%%%%%%%%%%%%%%%%%%%% 2HDM or MSSM plus singlet %%%%%%%%%%%%%%%%%%%%%%




%\cite{Alanne:2016wtx}
\bibitem{Alanne:2016wtx} 
  T.~Alanne, K.~Kainulainen, K.~Tuominen and V.~Vaskonen,
  ``Baryogenesis in the two doublet and inert singlet extension of the Standard Model,''
  arXiv:1607.03303 [hep-ph].
  %%CITATION = ARXIV:1607.03303;%%
  %1 citations counted in INSPIRE as of 22 Aug 2016

%\cite{Huber:2006wf}
\bibitem{Huber:2006wf} 
  S.~J.~Huber, T.~Konstandin, T.~Prokopec and M.~G.~Schmidt,
  ``Electroweak Phase Transition and Baryogenesis in the nMSSM,''
  Nucl.\ Phys.\ B {\bf 757}, 172 (2006)
  doi:10.1016/j.nuclphysb.2006.09.003
  [hep-ph/0606298].
  %%CITATION = doi:10.1016/j.nuclphysb.2006.09.003;%%
  %93 citations counted in INSPIRE as of 22 Aug 2016

%\cite{Cheung:2012pg}
\bibitem{Cheung:2012pg} 
  K.~Cheung, T.~J.~Hou, J.~S.~Lee and E.~Senaha,
  ``Singlino-driven Electroweak Baryogenesis in the Next-to-MSSM,''
  Phys.\ Lett.\ B {\bf 710}, 188 (2012)
  doi:10.1016/j.physletb.2012.02.070
  [arXiv:1201.3781 [hep-ph]].
  %%CITATION = doi:10.1016/j.physletb.2012.02.070;%%
  %15 citations counted in INSPIRE as of 03 Dec 2016

%\cite{Huang:2014ifa}
\bibitem{Huang:2014ifa} 
  W.~Huang, Z.~Kang, J.~Shu, P.~Wu and J.~M.~Yang,
  ``New insights in the electroweak phase transition in the NMSSM,''
  Phys.\ Rev.\ D {\bf 91}, no. 2, 025006 (2015)
  doi:10.1103/PhysRevD.91.025006
  [arXiv:1405.1152 [hep-ph]].
  %%CITATION = doi:10.1103/PhysRevD.91.025006;%%
  %17 citations counted in INSPIRE as of 22 Aug 2016

%\cite{Demidov:2016wcv}
\bibitem{Demidov:2016wcv} 
  S.~V.~Demidov, D.~S.~Gorbunov and D.~V.~Kirpichnikov,
  ``Split NMSSM with electroweak baryogenesis,''
  JHEP {\bf 1611}, 148 (2016)
  doi:10.1007/JHEP11(2016)148
  [arXiv:1608.01985 [hep-ph]].
  %%CITATION = doi:10.1007/JHEP11(2016)148;%%

%%%%%%%%%%%%%%%%%%%%%%% tree-level barrier %%%%%%%%%%%%%%%%%%%%%%%%%%%%%%%%%%%%%%%

%\cite{Choi:1993cv}
\bibitem{Choi:1993cv} 
  J.~Choi and R.~R.~Volkas,
  ``Real Higgs singlet and the electroweak phase transition in the Standard Model,''
  Phys.\ Lett.\ B {\bf 317}, 385 (1993)
  doi:10.1016/0370-2693(93)91013-D
  [hep-ph/9308234].
  %%CITATION = doi:10.1016/0370-2693(93)91013-D;%%
  %47 citations counted in INSPIRE as of 09 Mar 2017

%\cite{Espinosa:2011ax,Espinosa:2011eu}
\bibitem{Espinosa:2011ax} 
  J.~R.~Espinosa, T.~Konstandin and F.~Riva,
  ``Strong Electroweak Phase Transitions in the Standard Model with a Singlet,''
  Nucl.\ Phys.\ B {\bf 854}, 592 (2012)
  [arXiv:1107.5441 [hep-ph]].
  %%CITATION = ARXIV:1107.5441;%%

%\cite{Espinosa:2011eu}
\bibitem{Espinosa:2011eu} 
  J.~R.~Espinosa, B.~Gripaios, T.~Konstandin and F.~Riva,
  ``Electroweak Baryogenesis in Non-minimal Composite Higgs Models,''
  JCAP {\bf 1201}, 012 (2012)
  [arXiv:1110.2876 [hep-ph]].
  %%CITATION = ARXIV:1110.2876;%%

%%%%%%%%%%% works using tree level barrier %%%%%%%%%%%%%%%%%%%%%%%%%%%%%%%%%%%%%

%\cite{Cline:2012hg,Fairbairn:2013uta,Li:2014wia,Alanne:2014bra,Jiang:2015cwa,Xiao:2015tja}
\bibitem{Cline:2012hg} 
  J.~M.~Cline and K.~Kainulainen,
  ``Electroweak baryogenesis and dark matter from a singlet Higgs,''
  JCAP {\bf 1301}, 012 (2013)
  doi:10.1088/1475-7516/2013/01/012
  [arXiv:1210.4196 [hep-ph]].
  %%CITATION = doi:10.1088/1475-7516/2013/01/012;%%
  %67 citations counted in INSPIRE as of 22 Aug 2016

%\cite{Fairbairn:2013uta}
\bibitem{Fairbairn:2013uta} 
  M.~Fairbairn and R.~Hogan,
  ``Singlet Fermionic Dark Matter and the Electroweak Phase Transition,''
  JHEP {\bf 1309}, 022 (2013)
  doi:10.1007/JHEP09(2013)022
  [arXiv:1305.3452 [hep-ph]].
  %%CITATION = doi:10.1007/JHEP09(2013)022;%%
  %30 citations counted in INSPIRE as of 22 Aug 2016

%\cite{Li:2014wia}
\bibitem{Li:2014wia} 
  T.~Li and Y.~F.~Zhou,
  ``Strongly first order phase transition in the singlet fermionic dark matter model after LUX,''
  JHEP {\bf 1407}, 006 (2014)
  doi:10.1007/JHEP07(2014)006
  [arXiv:1402.3087 [hep-ph]].
  %%CITATION = doi:10.1007/JHEP07(2014)006;%%
  %14 citations counted in INSPIRE as of 22 Aug 2016

%\cite{Alanne:2014bra}
\bibitem{Alanne:2014bra} 
  T.~Alanne, K.~Tuominen and V.~Vaskonen,
  ``Strong phase transition, dark matter and vacuum stability from simple hidden sectors,''
  Nucl.\ Phys.\ B {\bf 889}, 692 (2014)
  doi:10.1016/j.nuclphysb.2014.11.001
  [arXiv:1407.0688 [hep-ph]].
  %%CITATION = doi:10.1016/j.nuclphysb.2014.11.001;%%
  %14 citations counted in INSPIRE as of 22 Aug 2016

%\cite{Jiang:2015cwa}
\bibitem{Jiang:2015cwa} 
  M.~Jiang, L.~Bian, W.~Huang and J.~Shu,
  ``Impact of a complex singlet: Electroweak baryogenesis and dark matter,''
  Phys.\ Rev.\ D {\bf 93}, no. 6, 065032 (2016)
  doi:10.1103/PhysRevD.93.065032
  [arXiv:1502.07574 [hep-ph]].
  %%CITATION = doi:10.1103/PhysRevD.93.065032;%%
  %12 citations counted in INSPIRE as of 22 Aug 2016

%\cite{Sannino:2015wka}
\bibitem{Sannino:2015wka} 
  F.~Sannino and J.~Virkajärvi,
  ``First Order Electroweak Phase Transition from (Non)Conformal Extensions of the Standard Model,''
  Phys.\ Rev.\ D {\bf 92}, no. 4, 045015 (2015)
  doi:10.1103/PhysRevD.92.045015
  [arXiv:1505.05872 [hep-ph]].
  %%CITATION = doi:10.1103/PhysRevD.92.045015;%%
  %8 citations counted in INSPIRE as of 09 Mar 2017

%\cite{Huang:2015bta}
\bibitem{Huang:2015bta} 
  F.~P.~Huang and C.~S.~Li,
  ``Electroweak baryogenesis in the framework of the effective field theory,''
  Phys.\ Rev.\ D {\bf 92}, no. 7, 075014 (2015)
  doi:10.1103/PhysRevD.92.075014
  [arXiv:1507.08168 [hep-ph]].
  %%CITATION = doi:10.1103/PhysRevD.92.075014;%%
  %6 citations counted in INSPIRE as of 09 Mar 2017

%\cite{Xiao:2015tja}
\bibitem{Xiao:2015tja} 
  M.~L.~Xiao and J.~H.~Yu,
  ``Electroweak baryogenesis in a scalar-assisted vectorlike fermion model,''
  Phys.\ Rev.\ D {\bf 94}, no. 1, 015011 (2016)
  doi:10.1103/PhysRevD.94.015011
  [arXiv:1509.02931 [hep-ph]].
  %%CITATION = doi:10.1103/PhysRevD.94.015011;%%
  %2 citations counted in INSPIRE as of 15 Aug 2016


%\cite{Vaskonen:2016yiu}
\bibitem{Vaskonen:2016yiu} 
  V.~Vaskonen,
  ``Electroweak baryogenesis and gravitational waves from a real scalar singlet,''
  arXiv:1611.02073 [hep-ph].
  %%CITATION = ARXIV:1611.02073;%%
  %2 citations counted in INSPIRE as of 03 Dec 2016

%%%%%%%%%%%%%%% EWBG + gravity waves %%%%%%%%%%%%%%%%%%%%%%%%%%%%%%%%%%%%

%\cite{Vaskonen:2016yiu,Dorsch:2016nrg,Artymowski:2016tme,Katz:2016adq,Chala:2016ykx,Huang:2016odd,No:2011fi,Dolgov:2000ht} 
%\cite{Artymowski:2016tme}
%\cite{No:2011fi}
\bibitem{No:2011fi}
  J.~M.~No,
  ``Large Gravitational Wave Background Signals in Electroweak Baryogenesis Scenarios,''
  Phys.\ Rev.\ D {\bf 84}, 124025 (2011)
  doi:10.1103/PhysRevD.84.124025
  [arXiv:1103.2159 [hep-ph]].
  %%CITATION = doi:10.1103/PhysRevD.84.124025;%%
  %22 citations counted in INSPIRE as of 04 Dec 2016
%\cite{Dolgov:2000ht}
\bibitem{Artymowski:2016tme}
  M.~Artymowski, M.~Lewicki and J.~D.~Wells,
  ``Gravitational wave and collider implications of electroweak baryogenesis aided by non-standard cosmology,''
  arXiv:1609.07143 [hep-ph].
  %%CITATION = ARXIV:1609.07143;%%
  %3 citations counted in INSPIRE as of 04 Dec 2016
%\cite{Katz:2016adq}
%\cite{Jaeckel:2016jlh}
\bibitem{Jaeckel:2016jlh} 
  J.~Jaeckel, V.~V.~Khoze and M.~Spannowsky,
  ``Hearing the signals of dark sectors with gravitational wave detectors,''
  Phys.\ Rev.\ D {\bf 94}, no. 10, 103519 (2016)
  doi:10.1103/PhysRevD.94.103519
  [arXiv:1602.03901 [hep-ph]].
  %%CITATION = doi:10.1103/PhysRevD.94.103519;%%
  %23 citations counted in INSPIRE as of 09 Mar 2017
\bibitem{Katz:2016adq}
  A.~Katz and A.~Riotto,
  ``Baryogenesis and Gravitational Waves from Runaway Bubble Collisions,''
  JCAP {\bf 1611}, no. 11, 011 (2016)
  doi:10.1088/1475-7516/2016/11/011
  [arXiv:1608.00583 [hep-ph]].
  %%CITATION = doi:10.1088/1475-7516/2016/11/011;%%
  %3 citations counted in INSPIRE as of 04 Dec 2016
%\cite{Chala:2016ykx}
\bibitem{Chala:2016ykx}
  M.~Chala, G.~Nardini and I.~Sobolev,
  ``Unified explanation for dark matter and electroweak baryogenesis with direct detection and gravitational wave signatures,''
  Phys.\ Rev.\ D {\bf 94}, no. 5, 055006 (2016)
  doi:10.1103/PhysRevD.94.055006
  [arXiv:1605.08663 [hep-ph]].
  %%CITATION = doi:10.1103/PhysRevD.94.055006;%%
  %3 citations counted in INSPIRE as of 04 Dec 2016
%\cite{Huang:2016odd}
\bibitem{Huang:2016odd}
  F.~P.~Huang, Y.~Wan, D.~G.~Wang, Y.~F.~Cai and X.~Zhang,
  ``Hearing the echoes of electroweak baryogenesis with gravitational wave detectors,''
  Phys.\ Rev.\ D {\bf 94}, no. 4, 041702 (2016)
  doi:10.1103/PhysRevD.94.041702
  [arXiv:1601.01640 [hep-ph]].
  %%CITATION = doi:10.1103/PhysRevD.94.041702;%%
  %10 citations counted in INSPIRE as of 04 Dec 2016
%\cite{Beniwal:2017eik}
\bibitem{Beniwal:2017eik} 
  A.~Beniwal, M.~Lewicki, J.~D.~Wells, M.~White and A.~G.~Williams,
  ``Gravitational wave, collider and dark matter signals from a scalar singlet electroweak baryogenesis,''
  arXiv:1702.06124 [hep-ph].
  %%CITATION = ARXIV:1702.06124;%%
\bibitem{Dolgov:2000ht} 
  A.~D.~Dolgov, P.~D.~Naselsky and I.~D.~Novikov,
  ``Gravitational waves, baryogenesis, and dark matter from primordial black holes,''
  %Submitted to: Phys.Rev.D
  [astro-ph/0009407].
  %%CITATION = ASTRO-PH/0009407;%%
  %33 citations counted in INSPIRE as of 04 Dec 2016


%%%%%%%%%%%%%% singlet models + EWPT +DM %%%%%%%%%%%%%%%%%%%%%%%%%%%%%%%%%%%%%%%%%%%%%%%%%%%%%%%%%

% \cite{McDonald:1993ey,Profumo:2007wc,Babu:2007sm,Barger:2008jx,Ahriche:2012ei,
% Gonderinger:2012rd,Fairbairn:2013xaa,Kanemura:2014cka,Jiang:2015cwa,
% Lewicki:2016efe,Chala:2016ykx}
\bibitem{McDonald:1993ey} 
  J.~McDonald,
  ``Electroweak baryogenesis and dark matter via a gauge singlet scalar,''
  Phys.\ Lett.\ B {\bf 323}, 339 (1994).
  %%CITATION = PHLTA,B323,339;%%

%\cite{Profumo:2007wc}
\bibitem{Profumo:2007wc} 
  S.~Profumo, M.~J.~Ramsey-Musolf and G.~Shaughnessy,
  ``Singlet Higgs phenomenology and the electroweak phase transition,''
  JHEP {\bf 0708}, 010 (2007)
  [arXiv:0705.2425 [hep-ph]].
  %%CITATION = ARXIV:0705.2425;%%

%\cite{Babu:2007sm}
\bibitem{Babu:2007sm} 
  K.~S.~Babu and E.~Ma,
  ``Singlet fermion dark matter and electroweak baryogenesis with radiative neutrino mass,''
  Int.\ J.\ Mod.\ Phys.\ A {\bf 23}, 1813 (2008)
  doi:10.1142/S0217751X08040299
  [arXiv:0708.3790 [hep-ph]].
  %%CITATION = doi:10.1142/S0217751X08040299;%%
  %35 citations counted in INSPIRE as of 03 Dec 2016


%\cite{Barger:2008jx}
\bibitem{Barger:2008jx} 
  V.~Barger, P.~Langacker, M.~McCaskey, M.~Ramsey-Musolf and G.~Shaughnessy,
  ``Complex Singlet Extension of the Standard Model,''
  Phys.\ Rev.\ D {\bf 79}, 015018 (2009)
  [arXiv:0811.0393 [hep-ph]].
  %%CITATION = ARXIV:0811.0393;%%

%\cite{Ahriche:2012ei}
\bibitem{Ahriche:2012ei} 
  A.~Ahriche and S.~Nasri,
  ``Light Dark Matter, Light Higgs and the Electroweak Phase Transition,''
  arXiv:1201.4614 [hep-ph].
  %%CITATION = ARXIV:1201.4614;%%

\bibitem{Gonderinger:2012rd} 
  M.~Gonderinger, H.~Lim and M.~J.~Ramsey-Musolf,
  ``Complex Scalar Singlet Dark Matter: Vacuum Stability and Phenomenology,''
  arXiv:1202.1316 [hep-ph].
  %%CITATION = ARXIV:1202.1316;%%
 
%\cite{Fairbairn:2013xaa}
\bibitem{Fairbairn:2013xaa} 
  M.~Fairbairn and P.~Grothaus,
  ``Baryogenesis and Dark Matter with Vector-like Fermions,''
  JHEP {\bf 1310}, 176 (2013)
  doi:10.1007/JHEP10(2013)176
  [arXiv:1307.8011 [hep-ph]].
  %%CITATION = doi:10.1007/JHEP10(2013)176;%%
  %10 citations counted in INSPIRE as of 15 Aug 2016

%\cite{Kanemura:2014cka}
\bibitem{Kanemura:2014cka} 
  S.~Kanemura, N.~Machida and T.~Shindou,
  ``Radiative neutrino mass, dark matter and electroweak baryogenesis from the supersymmetric gauge theory with confinement,''
  Phys.\ Lett.\ B {\bf 738}, 178 (2014)
  doi:10.1016/j.physletb.2014.09.013
  [arXiv:1405.5834 [hep-ph]].
  %%CITATION = doi:10.1016/j.physletb.2014.09.013;%%
  %12 citations counted in INSPIRE as of 03 Dec 2016

%\cite{Jiang:2015cwa}
%\bibitem{Jiang:2015cwa} 
%  M.~Jiang, L.~Bian, W.~Huang and J.~Shu,
%  ``Impact of a complex singlet: Electroweak baryogenesis and dark matter,''
%  Phys.\ Rev.\ D {\bf 93}, no. 6, 065032 (2016)
%  doi:10.1103/PhysRevD.93.065032
%  [arXiv:1502.07574 [hep-ph]].
%  %%CITATION = doi:10.1103/PhysRevD.93.065032;%%
%  %13 citations counted in INSPIRE as of 03 Dec 2016


\bibitem{Lewicki:2016efe} 
  M.~Lewicki, T.~Rindler-Daller and J.~D.~Wells,
  ``Enabling Electroweak Baryogenesis through Dark Matter,''
  JHEP {\bf 1606}, 055 (2016)
  doi:10.1007/JHEP06(2016)055
  [arXiv:1601.01681 [hep-ph]].
  %%CITATION = doi:10.1007/JHEP06(2016)055;%%
  %7 citations counted in INSPIRE as of 03 Dec 2016

%\cite{Chala:2016ykx}
%\bibitem{Chala:2016ykx} 
%  M.~Chala, G.~Nardini and I.~Sobolev,
%  ``Unified explanation for dark matter and electroweak baryogenesis with direct detection and gravitational wave signatures,''
%  Phys.\ Rev.\ D {\bf 94}, no. 5, 055006 (2016)
%  doi:10.1103/PhysRevD.94.055006
%  [arXiv:1605.08663 [hep-ph]].
%  %%CITATION = doi:10.1103/PhysRevD.94.055006;%%
%  %3 citations counted in INSPIRE as of 03 Dec 2016



%%%%%%%%%%%%%%%%%%%%%%%%%%%%%%%%%%%%%%%%%%%%%%%%%%%%%%%%%%%%%%%%%%%%%%%%%%%%%

%\cite{Coleman:1977py}
\bibitem{Coleman:1977py} 
  S.~R.~Coleman,
  ``The Fate of the False Vacuum. 1. Semiclassical Theory,''
  Phys.\ Rev.\ D {\bf 15}, 2929 (1977)
  Erratum: [Phys.\ Rev.\ D {\bf 16}, 1248 (1977)].
  doi:10.1103/PhysRevD.15.2929, 10.1103/PhysRevD.16.1248
  %%CITATION = doi:10.1103/PhysRevD.15.2929, 10.1103/PhysRevD.16.1248;%%
  %1678 citations counted in INSPIRE as of 21 Apr 2017

%\cite{Wainwright:2011kj}
\bibitem{Wainwright:2011kj} 
  C.~L.~Wainwright,
  ``CosmoTransitions: Computing Cosmological Phase Transition Temperatures and Bubble Profiles with Multiple Fields,''
  Comput.\ Phys.\ Commun.\  {\bf 183}, 2006 (2012)
  doi:10.1016/j.cpc.2012.04.004
  [arXiv:1109.4189 [hep-ph]].
  %%CITATION = doi:10.1016/j.cpc.2012.04.004;%%
  %80 citations counted in INSPIRE as of 21 Apr 2017


%\cite{Akula:2016gpl}
\bibitem{Akula:2016gpl} 
  S.~Akula, C.~Bal\'azs and G.~A.~White,
  ``Semi-analytic techniques for calculating bubble wall profiles,''
  Eur.\ Phys.\ J.\ C {\bf 76}, no. 12, 681 (2016)
  doi:10.1140/epjc/s10052-016-4519-5
  [arXiv:1608.00008 [hep-ph]].
  %%CITATION = doi:10.1140/epjc/s10052-016-4519-5;%%
  %2 citations counted in INSPIRE as of 24 Feb 2017

%\cite{Moore:1998swa}
\bibitem{Moore:1998swa} 
  G.~D.~Moore,
  ``Measuring the broken phase sphaleron rate nonperturbatively,''
  Phys.\ Rev.\ D {\bf 59}, 014503 (1999)
  doi:10.1103/PhysRevD.59.014503
  [hep-ph/9805264].
  %%CITATION = doi:10.1103/PhysRevD.59.014503;%%
  %93 citations counted in INSPIRE as of 25 Aug 2016

%\cite{Fuyuto:2014yia}
\bibitem{Fuyuto:2014yia} 
  K.~Fuyuto and E.~Senaha,
  ``Improved sphaleron decoupling condition and the Higgs coupling constants in the real singlet-extended standard model,''
  Phys.\ Rev.\ D {\bf 90}, no. 1, 015015 (2014)
  doi:10.1103/PhysRevD.90.015015
  [arXiv:1406.0433 [hep-ph]].
  %%CITATION = doi:10.1103/PhysRevD.90.015015;%%
  %31 citations counted in INSPIRE as of 01 Mar 2017


%\cite{Cline:2000nw}
\bibitem{Cline:2000nw} 
  J.~M.~Cline, M.~Joyce and K.~Kainulainen,
  ``Supersymmetric electroweak baryogenesis,''
  JHEP {\bf 0007}, 018 (2000)
  doi:10.1088/1126-6708/2000/07/018
  [hep-ph/0006119].
  %%CITATION = doi:10.1088/1126-6708/2000/07/018;%%
  %171 citations counted in INSPIRE as of 17 Jul 2016

%\cite{Fromme:2006wx}
\bibitem{Fromme:2006wx} 
  L.~Fromme and S.~J.~Huber,
  ``Top transport in electroweak baryogenesis,''
  JHEP {\bf 0703}, 049 (2007)
  [hep-ph/0604159].
  %%CITATION = HEP-PH/0604159;%%


%\cite{Kainulainen:2001cn}
\bibitem{Kainulainen:2001cn}
  K.~Kainulainen, T.~Prokopec, M.~G.~Schmidt and S.~Weinstock,
  ``First principle derivation of semiclassical force for electroweak baryogenesis,''
  JHEP {\bf 0106} (2001) 031
  doi:10.1088/1126-6708/2001/06/031
  [hep-ph/0105295].
  %%CITATION = doi:10.1088/1126-6708/2001/06/031;%%
  %76 citations counted in INSPIRE as of 21 Feb 2017


%\cite{Kainulainen:2002th}
\bibitem{Kainulainen:2002th}
  K.~Kainulainen, T.~Prokopec, M.~G.~Schmidt and S.~Weinstock,
  ``Semiclassical force for electroweak baryogenesis: Three-dimensional derivation,''
  Phys.\ Rev.\ D {\bf 66} (2002) 043502
  doi:10.1103/PhysRevD.66.043502
  [hep-ph/0202177].
  %%CITATION = doi:10.1103/PhysRevD.66.043502;%%
  %50 citations counted in INSPIRE as of 21 Feb 2017


%%%%%%%%%%%%%%%%%%%% wall velocity %%%%%%%%%%%%%%%%%%%%%%%%%%%%%%%%%%%%%%%%%%

%\cite{Huber:2013kj,Konstandin:2014zta}
\bibitem{Huber:2013kj} 
  S.~J.~Huber and M.~Sopena,
  ``An efficient approach to electroweak bubble velocities,''
  arXiv:1302.1044 [hep-ph].
  %%CITATION = ARXIV:1302.1044;%%
  %16 citations counted in INSPIRE as of 22 Aug 2016

%\cite{Konstandin:2014zta}
\bibitem{Konstandin:2014zta} 
  T.~Konstandin, G.~Nardini and I.~Rues,
  ``From Boltzmann equations to steady wall velocities,''
  JCAP {\bf 1409}, no. 09, 028 (2014)
  doi:10.1088/1475-7516/2014/09/028
  [arXiv:1407.3132 [hep-ph]].
  %%CITATION = doi:10.1088/1475-7516/2014/09/028;%%
  %10 citations counted in INSPIRE as of 22 Aug 2016

%\cite{Kozaczuk:2015owa}
\bibitem{Kozaczuk:2015owa} 
  J.~Kozaczuk,
  ``Bubble Expansion and the Viability of Singlet-Driven Electroweak Baryogenesis,''
  JHEP {\bf 1510}, 135 (2015)
  doi:10.1007/JHEP10(2015)135
  [arXiv:1506.04741 [hep-ph]].
  %%CITATION = doi:10.1007/JHEP10(2015)135;%%
  %18 citations counted in INSPIRE as of 22 Aug 2016

%%%%%%%%%%%%%%%%%%%%%%%%%%%%%%%%%%%%%%%%%%%%%%%%%%%%%%%%%%%%%%%%%%%%%%%%%%%%%%%%%%%%%%%%%






%\cite{Aad:2014yka}
\bibitem{Aad:2014yka} 
  G.~Aad {\it et al.} [ATLAS Collaboration],
  ``Search for the direct production of charginos, neutralinos and staus in final states with at least two hadronically decaying taus and missing transverse momentum in $pp$ collisions at $\sqrt{s}$ = 8 TeV with the ATLAS detector,''
  JHEP {\bf 1410}, 096 (2014)
  doi:10.1007/JHEP10(2014)096
  [arXiv:1407.0350 [hep-ex]].
  %%CITATION = doi:10.1007/JHEP10(2014)096;%%
  %74 citations counted in INSPIRE as of 10 Oct 2016
  
  %\cite{Khachatryan:2016trj}
\bibitem{Khachatryan:2016trj} 
  V.~Khachatryan {\it et al.} [CMS Collaboration],
  ``Search for electroweak production of charginos in final states with two tau leptons in pp collisions at sqrt(s) = 8 TeV,''
  %Submitted to: JHEP
  [arXiv:1610.04870 [hep-ex]].
  %%CITATION = ARXIV:1610.04870;%%
  %1 citations counted in INSPIRE as of 28 Feb 2017

%\cite{ATLAS:2016ety}
\bibitem{ATLAS:2016ety} 
  The ATLAS collaboration [ATLAS Collaboration],
  ``Search for electroweak production of supersymmetric particles in final states with tau leptons in $\sqrt{s} =$ 13TeV pp collisions with the ATLAS detector,''
  ATLAS-CONF-2016-093.
  %%CITATION = ATLAS-CONF-2016-093;%%
  %2 citations counted in INSPIRE as of 24 Feb 2017

%\cite{ATLAS:2016uwq}
\bibitem{ATLAS:2016uwq} 
  The ATLAS collaboration [ATLAS Collaboration],
  ``Search for supersymmetry with two and three leptons and missing
transverse momentum in the final state at $\sqrt{s}=13$ TeV with the ATLAS detector,''
  ATLAS-CONF-2016-096.
  %%CITATION = ATLAS-CONF-2016-096;%%
  %6 citations counted in INSPIRE as of 24 Feb 2017
  
\bibitem{LEP} 
http://lepsusy.web.cern.ch/lepsusy/www/sleptons\_
summer04/slep\_final.html; ALEPH, Phys. Lett. B526 (2002) 206;
DELPHI, Eur. Phys. J. C31 (2003) 421-479; 
L3, Phys. lett  B580 (2004) 37; OPAL, Eur. Phys. J C32 (2004) 453-473



%\cite{Kolb:1990vq}
\bibitem{Kolb:1990vq} 
  E.~W.~Kolb and M.~S.~Turner,
  ``The Early Universe,''
  Front.\ Phys.\  {\bf 69}, 1 (1990).
  %%CITATION = FRPHA,69,1;%%
  %1089 citations counted in INSPIRE as of 06 Dec 2016

%\cite{Yang:2016odq}
\bibitem{Yang:2016odq} 
  Y.~Yang [On behalf of PandaX-II Collaboration],
  ``Search for dark matter from the first data of the PandaX-II experiment,''
  arXiv:1612.01223 [hep-ex].
  %%CITATION = ARXIV:1612.01223;%%

%\cite{Chao:2016lqd}
\bibitem{Chao:2016lqd} 
  W.~Chao, H.~K.~Guo and H.~L.~Li,
  ``Tau flavored dark matter and its impact on tau Yukawa coupling,''
  JCAP {\bf 1702}, no. 02, 002 (2017)
  doi:10.1088/1475-7516/2017/02/002
  [arXiv:1606.07174 [hep-ph]].
  %%CITATION = doi:10.1088/1475-7516/2017/02/002;%%

%\cite{Ho:2012bg}
\bibitem{Ho:2012bg} 
  C.~M.~Ho and R.~J.~Scherrer,
  ``Anapole Dark Matter,''
  Phys.\ Lett.\ B {\bf 722}, 341 (2013)
  doi:10.1016/j.physletb.2013.04.039
  [arXiv:1211.0503 [hep-ph]].
  %%CITATION = doi:10.1016/j.physletb.2013.04.039;%%
  %32 citations counted in INSPIRE as of 08 Sep 2016

%\cite{Ackermann:2015lka}
\bibitem{Ackermann:2015lka} 
  M.~Ackermann {\it et al.} [Fermi-LAT Collaboration],
  ``Updated search for spectral lines from Galactic dark matter interactions with pass 8 data from the Fermi Large Area Telescope,''
  Phys.\ Rev.\ D {\bf 91}, no. 12, 122002 (2015)
  doi:10.1103/PhysRevD.91.122002
  [arXiv:1506.00013 [astro-ph.HE]].
  %%CITATION = doi:10.1103/PhysRevD.91.122002;%%
  %115 citations counted in INSPIRE as of 10 Apr 2017

%\cite{Garcia-Cely:2016hsk}
\bibitem{Garcia-Cely:2016hsk} 
  C.~Garcia-Cely and A.~Rivera,
  ``General calculation of the cross section for dark matter annihilations into two photons,''
  JCAP {\bf 1703}, no. 03, 054 (2017)
  doi:10.1088/1475-7516/2017/03/054
  [arXiv:1611.08029 [hep-ph]].
  %%CITATION = doi:10.1088/1475-7516/2017/03/054;%%
  %1 citations counted in INSPIRE as of 10 Apr 2017

%\cite{Bringmann:2007nk}
\bibitem{Bringmann:2007nk} 
  T.~Bringmann, L.~Bergstrom and J.~Edsjo,
  ``New Gamma-Ray Contributions to Supersymmetric Dark Matter Annihilation,''
  JHEP {\bf 0801}, 049 (2008)
  doi:10.1088/1126-6708/2008/01/049
  [arXiv:0710.3169 [hep-ph]].
  %%CITATION = doi:10.1088/1126-6708/2008/01/049;%%
  %267 citations counted in INSPIRE as of 10 Apr 2017

%\cite{Ackermann:2015zua}
\bibitem{Ackermann:2015zua} 
  M.~Ackermann {\it et al.} [Fermi-LAT Collaboration],
  ``Searching for Dark Matter Annihilation from Milky Way Dwarf Spheroidal Galaxies with Six Years of Fermi Large Area Telescope Data,''
  Phys.\ Rev.\ Lett.\  {\bf 115}, no. 23, 231301 (2015)
  doi:10.1103/PhysRevLett.115.231301
  [arXiv:1503.02641 [astro-ph.HE]].
  %%CITATION = doi:10.1103/PhysRevLett.115.231301;%%
  %400 citations counted in INSPIRE as of 10 Apr 2017

%\cite{Moore:2000wx}
\bibitem{Moore:2000wx}
  G.~D.~Moore,
  %``Electroweak bubble wall friction: Analytic results,''
  JHEP {\bf 0003} (2000) 006
  doi:10.1088/1126-6708/2000/03/006
  [hep-ph/0001274].
  %%CITATION = doi:10.1088/1126-6708/2000/03/006;%%
  %77 citations counted in INSPIRE as of 10 Apr 2017


%\cite{Notzold:1987ik}
\bibitem{Notzold:1987ik} 
  D.~Notzold and G.~Raffelt,
  ``Neutrino Dispersion at Finite Temperature and Density,''
  Nucl.\ Phys.\ B {\bf 307}, 924 (1988).
  doi:10.1016/0550-3213(88)90113-7
  %%CITATION = doi:10.1016/0550-3213(88)90113-7;%%
  %385 citations counted in INSPIRE as of 06 Sep 2016


%\cite{Kolb:1979qa,Cline:1993bd,Giudice:2003jh}
\bibitem{Kolb:1979qa} 
  E.~W.~Kolb and S.~Wolfram,
  ``Baryon Number Generation in the Early Universe,''
  Nucl.\ Phys.\ B {\bf 172}, 224 (1980)
  Erratum: [Nucl.\ Phys.\ B {\bf 195}, 542 (1982)].
  doi:10.1016/0550-3213(80)90167-4, 10.1016/0550-3213(82)90012-8
  %%CITATION = doi:10.1016/0550-3213(80)90167-4, 10.1016/0550-3213(82)90012-8;%%
  %311 citations counted in INSPIRE as of 19 Nov 2016

%\cite{Cline:1993bd}
\bibitem{Cline:1993bd} 
  J.~M.~Cline, K.~Kainulainen and K.~A.~Olive,
  ``Protecting the primordial baryon asymmetry from erasure by sphalerons,''
  Phys.\ Rev.\ D {\bf 49}, 6394 (1994)
  doi:10.1103/PhysRevD.49.6394
  [hep-ph/9401208].
  %%CITATION = doi:10.1103/PhysRevD.49.6394;%%
  %119 citations counted in INSPIRE as of 19 Nov 2016

%\cite{Giudice:2003jh}
\bibitem{Giudice:2003jh} 
  G.~F.~Giudice, A.~Notari, M.~Raidal, A.~Riotto and A.~Strumia,
  ``Towards a complete theory of thermal leptogenesis in the SM and MSSM,''
  Nucl.\ Phys.\ B {\bf 685}, 89 (2004)
  doi:10.1016/j.nuclphysb.2004.02.019
  [hep-ph/0310123].
  %%CITATION = doi:10.1016/j.nuclphysb.2004.02.019;%%
  %590 citations counted in INSPIRE as of 19 Nov 2016

%\cite{Weldon:1982bn}
%\bibitem{Weldon:1982bn} 
%  H.~A.~Weldon,
%  ``Effective Fermion Masses of Order gT in High Temperature Gauge Theories with Exact Chiral Invariance,''
%  Phys.\ Rev.\ D {\bf 26}, 2789 (1982).
%  doi:10.1103/PhysRevD.26.2789
%  %%CITATION = doi:10.1103/PhysRevD.26.2789;%%
%  %475 citations counted in INSPIRE as of 19 Nov 2016

%\cite{Shtabovenko:2016sxi}
\bibitem{Shtabovenko:2016sxi} 
  V.~Shtabovenko, R.~Mertig and F.~Orellana,
  ``New Developments in FeynCalc 9.0,''
  Comput.\ Phys.\ Commun.\  {\bf 207}, 432 (2016)
  doi:10.1016/j.cpc.2016.06.008
  [arXiv:1601.01167 [hep-ph]].
  %%CITATION = doi:10.1016/j.cpc.2016.06.008;%%
  %45 citations counted in INSPIRE as of 28 Feb 2017


\end{thebibliography}

\end{document}